\documentclass[twocolumn,showpacs,aps,prd,superscriptaddress]{revtex4}

\usepackage{graphicx}
\usepackage{dcolumn}
\usepackage{amsmath}
\usepackage{amssymb}
\usepackage{epsfig}
\usepackage{bm}
\newcommand{\Rab}{R_{12}}
\newcommand{\Rbc}{R_{23}}
\newcommand{\Rac}{R_{13}}
\newcommand{\Wab}{W_{12}}
\newcommand{\Wbc}{W_{23}}
\newcommand{\Wac}{W_{13}}

\newcommand{\RepA}{\mathcal{A}}
\newcommand{\RepB}{\mathcal{B}}
\newcommand{\RepC}{\mathcal{C}}
\newcommand{\RepD}{\mathcal{D}}
\newcommand{\RepE}{\mathcal{E}}
\newcommand{\RepF}{\mathcal{F}}
\newcommand{\RepG}{\mathcal{G}}
\newcommand{\RepH}{\mathcal{H}}
\newcommand{\RepS}{\mathcal{S}}

\newcommand{\CPV}{\delta_{cp}}
\newcommand{\CPVA}{\delta_{cp}^{\mathcal{A}} }
\newcommand{\CPVB}{\delta_{cp}^{\mathcal{B}} }
\newcommand{\CPVC}{\delta_{cp}^{\mathcal{C}} }
\newcommand{\CPVD}{\delta_{cp}^{\mathcal{D}} }
\newcommand{\CPVE}{\delta_{cp}^{\mathcal{E}} }
\newcommand{\CPVF}{\delta_{cp}^{\mathcal{F}} }
\newcommand{\CPVG}{\delta_{cp}^{\mathcal{G}} }
\newcommand{\CPVH}{\delta_{cp}^{\mathcal{H}} }
\newcommand{\CPVS}{\delta_{cp}^{\mathcal{S}} }

\newcommand{\thA}{\theta^{\mathcal{A}} }
\newcommand{\thB}{\theta^{\mathcal{B}} }
\newcommand{\thC}{\theta^{\mathcal{C}} }
\newcommand{\thD}{\theta^{\mathcal{D}} }
\newcommand{\thE}{\theta^{\mathcal{E}} }
\newcommand{\thF}{\theta^{\mathcal{F}} }
\newcommand{\thG}{\theta^{\mathcal{G}} }
\newcommand{\thH}{\theta^{\mathcal{H}} }
\newcommand{\thS}{\theta^{\mathcal{S}} }

\newcommand{\Jp}{\mathcal{J'} }
\newcommand{\JpS}{\mathcal{J}^{\mathcal{'S}} }

\newcommand{\PhiB}{\Phi^{\mathcal{B}} }
\newcommand{\PhiC}{\Phi^{\mathcal{C}} }
\newcommand{\PhiD}{\Phi^{\mathcal{D}} }
\newcommand{\PhiE}{\Phi^{\mathcal{E}} }
\newcommand{\PhiF}{\Phi^{\mathcal{F}} }

\newcommand{\etaB}{\eta^{\mathcal{B}} }
\newcommand{\etaC}{\eta^{\mathcal{C}} }

\newcommand{\etaE}{\eta^{\mathcal{E}} }
\newcommand{\etaF}{\eta^{\mathcal{F}} }

\newcommand{\xiB}{\xi^{\mathcal{B}} }

\newcommand{\cA}{c^{\mathcal{A}} }
\newcommand{\cB}{c^{\mathcal{B}} }
\newcommand{\cC}{c^{\mathcal{C}} }
\newcommand{\cD}{c^{\mathcal{D}} }
\newcommand{\cE}{c^{\mathcal{E}} }
\newcommand{\cF}{c^{\mathcal{F}} }

\newcommand{\sA}{s^{\mathcal{A}} }
\newcommand{\sB}{s^{\mathcal{B}} }
\newcommand{\sC}{s^{\mathcal{C}} }
\newcommand{\sD}{s^{\mathcal{D}} }
\newcommand{\sE}{s^{\mathcal{E}} }
\newcommand{\sF}{s^{\mathcal{F}} }

\newcommand{\ProbA}{P^{\mathcal{A}} }
\newcommand{\ProbB}{P^{\mathcal{B}} }
\newcommand{\ProbC}{P^{\mathcal{C}} }
\newcommand{\ProbD}{P^{\mathcal{D}} }
\newcommand{\ProbE}{P^{\mathcal{E}} }
\newcommand{\ProbF}{P^{\mathcal{F}} }

\begin{document}

% ---------- Title ----------- %
%\input{title_abs_NuParam_v6.tex}

\title{
\begin{center}
{\Large \textbf{
   Dependence of Neutrino Mixing Angles and CP-violating \vspace{0.2cm}
              Phase on Mixing Matrix Parametrizations  }}
\end{center}
}

\affiliation{Leung Center for Cosmology and Particle Astrophysics,
             National Taiwan University, Taipei 10617, Taiwan}
\affiliation{Department of Physics, Laurentian University, Sudbuary,
             ON P3E 2C6, Canada}
\affiliation{Department of Physics, University of Illinois,
             Urbana-Champaign,  IL 61801, USA}
\affiliation{Institute of Physics, Academia Sinica, Taipei 11529, Taiwan}
\affiliation{Fermi National Accelerator Laboratory, Batavia, IL 60510, USA}
\affiliation{Department of Physics, University of Guelph, Guelph,
             ON N1G 2W1, Canada}
\affiliation{Graduate Institute of Physics, National Taiwan University,
             Taipei 10617, Taiwan}

\author{Melin Huang}
%\thanks{Corresponding author, email:phmelin@phys.ntu.edu.tw}
   \affiliation{Leung Center for Cosmology and Particle Astrophysics,
                National Taiwan University, Taipei 10617, Taiwan}
   \affiliation{Department of Physics, Laurentian University, Sudbuary,
                ON P3E 2C6, Canada}
\author{Dawei Liu}
   \affiliation{Department of Physics, University of Illinois,
                Urbana-Champaign, IL 61801, USA}
\author{Jen-Chieh Peng}
   \affiliation{Department of Physics, University of Illinois,
                Urbana-Champaign, IL 61801, USA}
\affiliation{Institute of Physics, Academia Sinica, Taipei 11529, Taiwan}
\author{S. D. Reitzner}
   \affiliation{Fermi National Accelerator Laboratory, Batavia, IL 60510, USA}
   \affiliation{Department of Physics, University of Guelph,
                Guelph, ON N1G 2W1, Canada}
\author{Wei-Chun Tsai}
   \affiliation{Graduate Institute of Physics, National Taiwan University,
             Taipei 10617, Taiwan}

\date{\today}

% ----------- Abstract ------------ %
\begin{abstract}
\vspace{0.1in}
We consider various neutrino mixing matrix parametrizations and the dependence
of the mixing angles and CP-violating phase on the different parametrizations.
The transformations of neutrino mixing parameters between various
parametrizations are presented. Although the $\theta_{13}$ mixing angle
is determined to be small in the conventional parametrization, we note that
in several other parametrizations the values of $\theta_{13}$ are quite
large. Should the value of $\theta_{13}$ turn out to be tiny in the
conventional parametrization, this study suggests that other alternative
mixing matrix representations would be more suitable for dertermining the
value of the CP-violating phase.
\end{abstract}


\maketitle

% ============= Section{Introduction} =============================== %
\section{Introduction}\label{sec:intro} 

Neutrino flavor oscillation has been well established by observations from
experiments involving solar \cite{Homestake_b8, Gallex_ga, GNO_ga, Sage_pp, 
Borexino, SK_II, SNO_2010}, reactor \cite{KamLAND_th13}, atmospheric
\cite{SK_atmos_I}, and accelerator \cite{K2K_02, MINOS_numu_dis} neutrinos.
Central for describing neutrino oscillation phenomenology is the
Pontecorvo-Maki-Nakagawa-Sakata (PMNS) mixing matrix \cite{PMNS_01, PMNS_03}.
The mixing matrix is an invariant quantity, but the parametrization of
the mixing matrix can be of different forms~\cite{Valle_Param, 
Fritzsch_Param, Zheng_Param, Giunti_Kim}.
The conventional parametrization for the mixing matrix of Dirac neutrinos
follows the convention adopted for the quark mixing, proposed in
1984~\cite{CKM_Param} prior to the observation of neutrino oscillation.
%%%%%%%%%%%%%%%%%%%%%%%%

For three active neutrinos with no sterile neutrino, the mixing matrix can
be expressed as a product of three rotation matrices. We define
%%%
\begin{equation*}
R_{23} = 
  \left(
     \begin{array}{ccc}
       1         & 0          & 0       \\
       0         &  c_{23}    & s_{23}  \\
       0         & -s_{23}    & c_{23} 
     \end{array}
  \right),  \hskip0.5cm
R_{13} = 
  \left(
     \begin{array}{ccc}
       c_{13}    & 0          & s_{13}   \\
       0         & 1          & 0        \\
      -s_{13}    & 0          & c_{13} 
     \end{array}
  \right),  \vspace{-0.3cm} 
\end{equation*}
\begin{equation}
R_{12} = 
  \left(
     \begin{array}{ccc}
       c_{12}    & s_{12}     & 0        \\
      -s_{12}    & c_{12}     & 0        \\
       0         & 0          & 1 
     \end{array}
  \right),   \vspace{-0.3cm}   
\label{eq:NuParam_01}
\end{equation}
and
\begin{equation*}
W_{23} =  
  \left(
    \begin{array}{ccc}
       1   &   0                  &   0                  \\
       0   &   c_{23}             &   s_{23} e^{-i\CPV}  \\
       0   &  -s_{23} e^{i\CPV}   &   c_{23} 
    \end{array}    
  \right),        \nonumber \\ 
\end{equation*}
\begin{equation*}
W_{13} =  
  \left(
    \begin{array}{ccc}
       c_{13}             &   0   &   s_{13} e^{-i\CPV}  \\
       0                  &   1   &   0                  \\
      -s_{13} e^{i\CPV}   &   0   &   c_{13} 
    \end{array}
  \right),         \nonumber \\ 
\end{equation*}
\begin{equation}
W_{12} =  
  \left(
    \begin{array}{ccc}
       c_{12}             &   s_{12} e^{-i\CPV}   &  0   \\
      -s_{12} e^{i\CPV}   &   c_{12}              &   0  \\
       0                  &   0                   &   1 
    \end{array}
  \right), 
\label{eq:NuParam_02}
\end{equation}
where $\theta_{ab}$ and $\CPV$ are the mixing angles and CP phase,
respectively, and
$c_{ab} \equiv \cos\theta_{ab}$ and $s_{ab} \equiv \sin\theta_{ab}$.
There are several different ways to place the $\CPV$ in the
rotation matrices.  We follow the
standard CKM mixing matrix for the $\CPV$ placement in this work.
%%%%%%%%%%%%%%%%%%%%%%%

The conventional ordering of the mixing matrix has been taken to be the
product of $\Rbc \times \Wac \times \Rab$. An important feature of such
a parametrization is that the three mixing angles are almost decoupled
for the solar, atmospheric, and reactor neutrino oscillation experiments.
In particular, the solar neutrino experiments are mostly sensitive to
$\theta_{12}$, the atmospheric neutrino experiments are more susceptible
to $\theta_{23}$, and the $\theta_{13}$ angle is probed in the
short-baseline reactor experiments.
Another interesting feature of the conventional parametrization is that
$\theta_{23} \sim 45^o$, corresponding to maximal mixing, while
$\theta_{13} \sim 0^o$, signifying minimal mixing. Since this convention
for parametrizing the neutrino mixing matrix was adopted prior to
the extraction of mixing angles from neutrino oscillation experiments,
it is interesting to examine whether or not these features would
be preserved in other possible parametrizations. In this paper, we
try to address several questions. First, how do the three mixing angles
depend on the parametrization of the mixing matrix? Will the three mixing
angles always be nicely decoupled for different oscillation experiments?
How are the $\CPV$ values in different parametrizations related? Finally,
is there any particular parametrization better suited for determining the
value of the $\CPV$ phase?
%%%%%%%%%%%%%%%%%%%%%%%

This paper is organized as follows. The parametrizations of the mixing matrix
are presented in Section~\ref{sec:NuParam_zero_CP} for the case of $\CPV = 0$
and in Section~\ref{sec:NuParam_nonzero_CP} for the non-zero $\CPV$ case,
respectively. The transformations of the three mixing angles and the one
CP-violating phase from one parametrization to another one are also presented 
in Section~\ref{sec:NuParam_nonzero_CP}.
A discussion on how the expressions for survival or transition probabilities
depend on the mixing matrix parametrization is given in
Section~\ref{sec:NuParam_Prob}. We then present the relevant expressions
for investigating $\CPV$ in various parametrizations in 
Section~\ref{sec:NuParam_CP_Asym}.
We show that certain parametrizations are more suitable for determining
the $\CPV$ phase if $\theta_{13}$ has a very small value in the conventional
parametrization.
A conclusion is given in Section~\ref{sec:NuParam_concl}.
%%%%%%%%%%%%%%%%%%%%%%%%%

% ========== Section{Mixing Parametrizations for $\CPV = 0$} ======== %
\section{\hskip-0.1cm Mixing Parametrizations for $\CPV = 0$}
     \label{sec:NuParam_zero_CP}
     %\input{NuParam_21_zero_CP_v6.tex}

% ------- Table{tbl_NuParam_zero_CP_ex1_v6.tex} ------- %
%
\begin{table*}[ht]
\begin{center}  %\vspace{0.4cm}
\begin{tabular}{ccccc||ccc} \hline\hline
 Rep. & $U$ Mixing Matrix
      & $\theta_{23}$ \hskip0.7cm  & $\theta_{13}$ \hskip0.7cm
      & $\theta_{12}$ \hskip0.7cm
      & $\theta_{23}$ \hskip0.7cm  & $\theta_{13}$ \hskip0.7cm
      & $\theta_{12}$ \hskip0.7cm                                \\ \hline
 $\RepA$
      &  $\Rbc \Rac \Rab$
      &  45.00 \hskip0.7cm  &  5.44  \hskip0.7cm  & 33.91
      &  45.00 \hskip0.7cm  &  0.00  \hskip0.7cm  & 33.91    \\
 $\RepB$
      &  $\Rbc \Rab \Rac$
      &  48.65 \hskip0.7cm  &  6.55  \hskip0.7cm  & 33.74
      &  45.00 \hskip0.7cm  &  0.00  \hskip0.7cm  & 33.91    \\
 $\RepC$
      &  $\Rac \Rbc \Rab$
      &  44.74 \hskip0.7cm  &  7.68  \hskip0.7cm  & 39.33
      &  45.00 \hskip0.7cm  &  0.00  \hskip0.7cm  & 33.91    \\
 $\RepD$
      &  $\Rac \Rab \Rbc$
      &  52.03 \hskip0.7cm  & -22.30 \hskip0.7cm  & 26.76
      &  50.31 \hskip0.7cm  & -25.43 \hskip0.7cm  & 23.24    \\
 $\RepE$
      &  $\Rab \Rbc \Rac$
      &  38.63 \hskip0.7cm  & -25.71 \hskip0.7cm  & 45.31
      &  35.93 \hskip0.7cm  & -29.16 \hskip0.7cm  & 43.55    \\
 $\RepF$
      &  $\Rab \Rac \Rbc$
      &  41.57 \hskip0.7cm  & -19.81 \hskip0.7cm  & 28.59
      &  39.69 \hskip0.7cm  & -23.24 \hskip0.7cm  & 25.43   \\  \hline\hline
\end{tabular}
\caption{Neutrino mixing angles in different representations for the case
         of $\CPV = 0$. }
\label{table:NuParam_zero_CP_ex} \vspace{-0.4cm}
\end{center}
\end{table*}
%
%
% --------- Contents --------- %
The state of a neutrino can be expressed either in the flavor eigenstate 
basis, $|\nu_{\alpha}>~(\alpha = e, \mu, \tau)$, or in the mass eigenstate 
basis, $|\nu_{k}>~(k = 1, 2, 3)$. The transformation from mass eigenstates 
to flavor eigenstates is described by a unitary mixing matrix $U$:
\begin{equation}
|\nu_{\alpha}> = \sum_{k} U_{\alpha k} |\nu_{k}>  \; .
\label{eq:NuParam_20}
\end{equation}
The parametrizations of $U$ can be acquired using three Euler angles
$(\theta_{12}, \theta_{23}, \theta_{13})$ (denoted as $\theta_{ij}$) and
one CP-violating phase, $\CPV$. One category of parametrizations for $U$
involves rotations around three distinct axes sequentially: $\Rbc \Wac \Rab$,
$\Rbc \Wab \Rac$, $\Rac \Wbc \Rab$, $\Rac \Wab \Rbc$, $\Rab \Wbc \Rac$,
$\Rab \Wac \Rbc$, where $\Rbc \Wac \Rab$ is the conventional parametrization.
In this paper, we will call these six different parametrizations as
$\RepA$--$\RepF$ `representations', respectively.
The other category is for rotations around two distinct axes: $\Rab \Wbc \Rab$,
$\Rab \Wac \Rab$, $\Rac \Wbc \Rac$, $\Rac \Wab \Rac$, $\Rbc \Wab \Rbc$,
$\Rbc \Wac \Rbc$.
It has been demonstrated that the second category can be reduced to only
three independent parametrizations \cite{Fritzsch_Param}.
We here limit our consideration to the rotations around three distinct axes
for mixing parametrizations.
%%%%%%%%%%%%%%%%%%%%%

Appendix~\ref{apdx:NuParam_UMatrix_zeroCP} presents the expressions of $U$
in terms of $\theta_{ij}$ in various representations for the special case
of $\CPV = 0$.  Given the elements of $U$, $U_{\alpha k}$, one can solve
$\theta_{ij}$ for any representation, also presented in this appendix. Note
that the values of $\theta_{ij}$ vary from one representation to another.
%%%%%%%%%%%%%%%%%%%%% 

Taking the central values of the three neutrino mixing angles from
\cite{KamLAND_th13} obtained from the conventional representation,
we present the values of the three mixing angles in other representations
in Table~\ref{table:NuParam_zero_CP_ex}. Again, the symbols $\RepA$--$\RepF$
denote various representations.
It can be seen that parametrizations
$\Rbc \Rac \Rab$, $\Rbc \Rab \Rac$, and $\Rac \Rbc \Rab$ produce small
values of $\theta_{13}$ while parametrizations
$\Rac \Rab \Rbc$, $\Rab \Rbc \Rac$, and $\Rab \Rac \Rbc$
generate significant non-zero values for $\theta_{13}$.
It is interesting to note that the existing neutrino oscillation data
favor a small or zero value for $\theta_{13}$ when the representations
$\RepA$, $\RepB$, $\RepC$ are chosen. In contrast, large non-zero central
values for $\theta_{13}$ would already have been obtained from existing
data if representations $\RepD$, $\RepE$, $\RepF$ were chosen.
The case with $\theta_{13} = 0$ and $\theta_{23} = 45^o$ in the conventional
representation is also shown in Table~\ref{table:NuParam_zero_CP_ex}.
Although the mixing angles in representations $\RepA$, $\RepB$, $\RepC$
are identical in this case, they are very different when representations
$\RepD$, $\RepE$, $\RepF$ are adopted.
%%%%%%%%%%%%%%%%%%%%%%%%%

% ======== Section{Mixing Parametrizations for $\CPV \ne 0$} ====== %
\section{\hskip-0.1cm Mixing Parametrizations for $\CPV \ne 0$}
      \label{sec:NuParam_nonzero_CP}
      %\input{NuParam_31_CP_0_v6.tex}

% ------- Contents -------- %
For the more general case of $\CPV \ne 0$, the transformations of
$(\theta_{ij}, \CPV)$ from one representation to another is more involved.
In the following, we show how these transformations can be obtained.

% ======== SubSection{Transformations between Representations} ====== %
\subsection{Transformations between Representations}
      \label{subsec:NuParam_nonzero_CP_1}
      %\input{NuParam_31_CP_1_v6.tex}

% ------- Contents -------- %
In general, the relation between two different representations, say 
$\RepG$ and $\RepH$ which can be any representation among $\RepA$--$\RepF$, 
can be expressed as
\begin{equation}
U = RWR(\thG_{ij}, \CPVG) = D^L \cdot RWR(\thH_{ij}, \CPVH) \cdot D^R.
\label{eq:NuParam_39_1}
\end{equation}
$D^L$ and $D^R$ are diagonal unitary matrices given as
\begin{eqnarray}
D^L &=& diag \left( e^{i\Psi_{L1}}, e^{i\Psi_{L2}}, e^{i\Psi_{L3}} 
                                                  \right)  \; , \nonumber \\
D^R &=& diag \left( e^{i\Psi_{R1}}, e^{i\Psi_{R2}}, e^{i\Psi_{R3}} 
                                                  \right)  \; , 
\label{eq:NuParam_39_2}
\end{eqnarray}
where $\Psi_{Li}$'s and $\Psi_{Rj}$'s are six phases.
Equations (\ref{eq:NuParam_39_1}) and (\ref{eq:NuParam_39_2}) show that
one can obtain the elements of $U$ in representation $\RepG$ by respectively
multiplying $\Psi_{Li}$ and $\Psi_{Rj}$ to the $i$-th row and $j$-th column
of the elements of $U$ in representations $\RepH$. In reality, there are only
five independent phases out of six.  This can be done by factoring out one
of the phases in $D^L$ or in $D^R$ and as a
result only five phases remain. For instance, one can factor out
$e^{i\Psi_{L2}}$ in $D^L$ and simultaneously multiply the factor
$e^{i\Psi_{L2}}$ in $D^R$. Consequently, $D^L$ and $D^R$ become
\begin{eqnarray*}
D^L &\Rightarrow& 
    diag \left( e^{i(\Psi_{L1} - \Psi_{L2})}, 1, 
                e^{i(\Psi_{L3} - \Psi_{L2})} \right)                 \\ 
    &\equiv& diag \left( e^{i\Phi_{L1}}, 1, e^{i\Phi_{L3}} \right)  \; ,  \\
D^R &\Rightarrow& 
    diag \left( e^{i(\Psi_{R1} + \Psi_{L2})}, 
                e^{i(\Psi_{R2} + \Psi_{L2})}, 
                e^{i(\Psi_{R3} + \Psi_{L2})} \right)                 \\
    &\equiv& diag \left( e^{i\Phi_{R1}}, e^{i\Phi_{R2}}, e^{i\Phi_{R3}} \right)
                                                                    \; .  
\end{eqnarray*}
This is equivalent to applying a set of five arbitrary phases
\begin{equation*}
\Phi = (\Phi_{L1}, 0, \Phi_{L3}, \Phi_{R1}, \Phi_{R2}, \Phi_{R3}) \; .
\end{equation*}
This suggests that only five independent phases are involved to transform
$(\thH_{ij}, \CPVH)$ to $(\thG_{ij}, \CPVG)$.
The solution for expressing $(\thG_{ij}, \CPVG)$ in
terms of $(\thH_{ij}, \CPVH)$ is unique and is independent of the
five phases. This will be illustrated in the next section using
transformation from representations $\RepA$ to $\RepD$ as an example.
%%%%%%%%%%%%%%%%%%%%%%%%%%%

% ========== SubSection{$(\thD_{ij}, \CPVD)$ Solutions} ========== %
\subsection{$(\thD_{ij}, \CPVD)$ Solutions}
      \label{subsec:NuParam_nonzero_CP_3}
      %\input{NuParam_31_CP_3_AD_v6.tex}

% -------- Contents ---------- %
Consider the case that $(\thA_{ij}, \CPVA)$ in the conventional representation 
$\RepA$ are known, and one would like to solve for $(\thD_{ij}, \CPVD)$ in 
representation $\RepD$. Equation (\ref{eq:NuParam_39_1}) becomes
\begin{equation}
U = RWR(\thD_{ij}, \CPVD) = D^L \cdot RWR(\thA_{ij}, \CPVA) \cdot D^R \; .
\label{eq:NonZeroCP_32_2}
\end{equation}
Take the $D^L$ and $D^R$ matrices as
\begin{eqnarray}
D^L
  &=& diag \left( e^{i\PhiD_{L1}}, e^{i\PhiD_{L2}}, e^{i\PhiD_{L3}} \right)  
                                                     \; ,  \nonumber \\
D^R
  &=& diag \left( e^{i\PhiD_{R1}}, e^{i\PhiD_{R2}}, 1  \right)   \; .  
\label{eq:NonZeroCP_31}
\end{eqnarray}
There are nine independent parameters in
Eq.(\ref{eq:NonZeroCP_32_2}), five phases in $D^L$ and $D^R$, three
mixing angles and one $\CPV$ phase from the mixing matrix.
The solutions for the nine parameters can be obtained
as follows (for derivation, see Appendix~\ref{apdx:NuParam_Solution_AD}).
\begin{equation}
\PhiD_{L1} + \PhiD_{R1} = 0  \hskip0.5cm \rm{and}  \hskip0.5cm
\PhiD_{L2} = 0 \; .
\label{eq:NonZeroCP_33_1}
\end{equation}
\begin{equation}
\sin^2{\PhiD_{R2}} = \frac{ (a~\sin\CPVA )^2}
                         { (a~\sin\CPVA )^2 + (a~\cos\CPVA - b)^2}  \; ,
\label{eq:NonZeroCP_33_2}
\end{equation}
where $a = \sA_{23} \sA_{12} \sA_{13}$ and $b = \cA_{23} \cA_{12}$ .
The sign of $\PhiD_{R2}$ can be determined through the equation
leading to Eq.(\ref{eq:NonZeroCP_33_2}).
(see Appendix~\ref{apdx:NuParam_Solution_AD}). We also have
\begin{equation}
\sin^2{\PhiD_{34}} = 
      \frac{ (a^{'} \sin\CPVA )^2 }
           { (a^{'} \sin\CPVA )^2 + (a^{'} \cos\CPVA - b^{'})^2 } \; ,
\label{eq:NonZeroCP_33_3}
\end{equation}
where $a^{'} = \cA_{23} \cA_{12} \sA_{13}$,  $b^{'} = \sA_{23} \sA_{12}$,
and $\PhiD_{34} \equiv \PhiD_{L3} + \PhiD_{R1}$.
Again, one can readily determine the sign of $\PhiD_{34}$ (see
Appendix~\ref{apdx:NuParam_Solution_AD}).
%%%%%%%%%%%%%%%%%%%%%%%%

After obtaining the phases $\PhiD_{R2}$ and $\PhiD_{34}$, the three mixing 
angles in representation $\RepD$ can be calculated:
\begin{eqnarray}
& & \hskip-1.2cm
  \tan \thD_{23} =
       \frac{\sA_{23} \cA_{13}}
            {\cA_{23} \cA_{12} \cos\PhiD_{R2} 
                - \sA_{23} \sA_{12}\sA_{13}\cos(\CPVA + \PhiD_{R2})} ,
\label{eq:NonZeroCP_33_4}
\end{eqnarray}
\begin{eqnarray}
& & \hskip-1.2cm
  \cos \thD_{12} = 
       \frac{\sA_{23} \cA_{13}}{\sD_{23}}  \nonumber \\
& & \hskip-0.1cm
   = \frac{\cA_{23}\cA_{12}\cos\PhiD_{R2} 
                - \sA_{23}\sA_{12}\sA_{13}\cos(\CPVA + \PhiD_{R2})}
          {\cD_{23}} , 
\label{eq:NonZeroCP_33_5}
\end{eqnarray}
\begin{eqnarray}
& & \hskip-1.2cm
  \sin \thD_{13} = 
      \frac{ \cA_{23}\cA_{12}\sA_{13}\cos(\CPVA + \PhiD_{34})
           - \sA_{23}\sA_{12}\cos\PhiD_{34}} {\cD_{12}}.
\label{eq:NonZeroCP_33_6}
\end{eqnarray}
The remaining parameters can be readily determined. First,
\begin{eqnarray}
& & \hskip-1.2cm
  \cos(\CPVA - \PhiD_{L1}) = 
       \frac{( \cD_{23}\sD_{13} )^2 + ( \sA_{13} )^2 
                                    - ( \sD_{23}\sD_{12}\cD_{13} )^2 }  
            {2 \cD_{23}\sD_{13} \sA_{13} }.
\label{eq:NonZeroCP_33_7}
\end{eqnarray}
Thus $\PhiD_{L1}$ can be calculated.  Subsequently,
$\PhiD_{R1} = - \PhiD_{L1}$ and $\PhiD_{L3} = \PhiD_{34} - \PhiD_{R1}$
can be obtained. Finally, the CP-violating phase in representation
$\RepD$, $\CPVD$, can be calculated using the expressions associated
with $\sin \CPVD$ in Eq.(\ref{eq:NonZeroCP_32_2}), listed as
follows:
\begin{eqnarray}
& & \hskip-1.0cm 
  \sin \CPVD                                \nonumber \\
& & \hskip-1.0cm
  = \frac{ \sA_{23}\cA_{12}\sA_{13} \sin(\CPVA + \eta) 
          +\cA_{23}\sA_{12} \sin \eta } { \sD_{12} } 
                                            \nonumber \\ 
& & \hskip-1.0cm
  = \frac{ \cA_{23}\sA_{12}\sA_{13} \sin(\CPVA + \xi) 
          +\sA_{23}\cA_{12} \sin \xi } {-\cD_{23}\sD_{12}\sD_{13} } 
                                            \nonumber \\ 
& & \hskip-1.0cm
  = \frac{-\sA_{12}\cA_{13} \sin(\PhiD_{L1} + \PhiD_{R2}) }  
         { \cD_{23}\sD_{12}\cD_{13} }        \nonumber \\
& & \hskip-1.0cm
  = \frac{\sA_{13} \sin(\CPVA - \PhiD_{L1}) }  
         {\sD_{23}\sD_{12}\cD_{13} }        \nonumber \\
& & \hskip-1.0cm
  = \frac{\cA_{23}\cA_{13} \sin \PhiD_{L3} }  
         {\sD_{23}\sD_{12}\sD_{13} }  \; ,  
\label{eq:NonZeroCP_33_8}
\end{eqnarray}
where $\eta \equiv \PhiD_{L2} + \PhiD_{R1}$  and
$\xi  \equiv \PhiD_{L3} + \PhiD_{R2}$.
Through these five different expressions of $\sin \CPVD$, the consistency
of the $\CPVD$ values can be checked.
%%%%%%%%%%%%%%%%%%%%%%%

With the expressions of $\PhiD$'s, one can further express $(\thD_{ij}, \CPVD)$
in terms of $(\thA_{ij}, \CPVA)$ only. As a result, the relations
between $(\thA_{ij}, \CPVA)$ and $(\thD_{ij}, \CPVD)$ are decoupled from
$\PhiD$'s. In other words, the matrices $D^L$ and $D^R$ only
act as a bridge in the transformation of $(\theta_{ij}, \CPV)$ from one
representation to another.
Note that $\CPV$ also plays a role in the transformation, as discussed
in the next section.
%%%%%%%%%%%%%%%%%%%%%%%

In Appendix~\ref{apdx:NuParam_Solution_0}, a detailed derivation of
transforming $(\thA_{ij}, \CPVA)$ to $(\thD_{ij}, \CPVD)$ is given
followed by listings of the solutions for the nine parameters used
in the transformations of $(\thA_{ij}, \CPVA)$ to $(\theta_{ij}, \CPV)$
in the remaining representations.
%%%%%%%%%%%%%%%%%%%%%%%

% ======== SubSection{Mixing Parameters in Neutrino Sector} ====== %
\subsection{Mixing Parameters in Neutrino Sector}
      \label{subsec:NuParam_nonzero_CP_4}
      %\input{NuParam_31_CP_4_neutrino_v6.tex}

\begin{figure}
\centerline{\psfig{figure=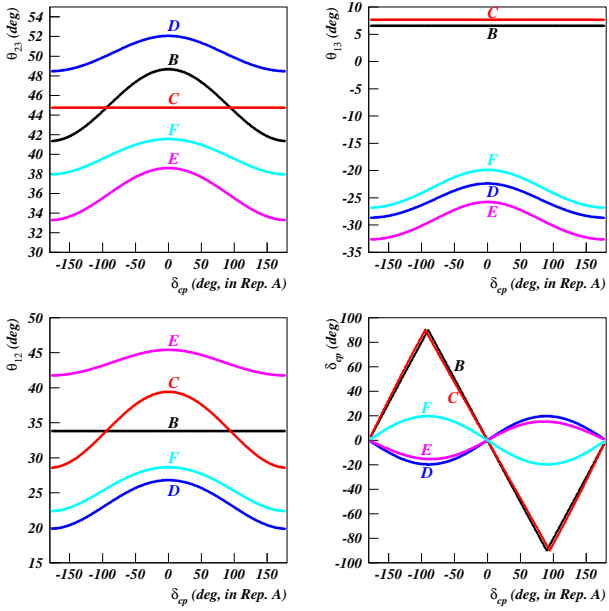,width=87mm}}
\caption{Values of ($\theta_{ij}$, $\CPV$) in different representations
         as a function of $\CPVA$, where $\thA_{12} = 34.0^o$,
         $\thA_{23} = 45^o$, and $\thA_{13} = 5.44^o$ are assumed;
         curves in different colors indicate different representations. }
\label{fig:Nuparam_Acp_BCDEF_th_cp}
\end{figure}

% -------- Contents ----------- %
Using the recent result of neutrino oscillation parameters \cite{KamLAND_th13}
along with $\thA_{23} = 45^o$, Fig.~\ref{fig:Nuparam_Acp_BCDEF_th_cp} shows
how $\CPVA$ influences ($\theta_{ij}$, $\CPV$) in other representations.
As can be seen from Eqs.(\ref{eq:Append_C_AB_6}) and
(\ref{eq:Append_C_AC_7}), $\thB_{13}$ and $\thC_{13}$ are independent of
$\CPVA$. Furthermore, $\thB_{12}$ and $\thC_{23}$ become independent of
$\CPVA$ after replacing the phases $\Phi$'s by $(\thA_{ij}, \CPVA)$ even though
they are functions of $\CPVA$ explicitly [see Eqs.(\ref{eq:Append_C_AB_5})
and (\ref{eq:Append_C_AC_6})].
It is interesting to note that the dependence of the new $\theta_{12}$,
$\theta_{13}$, and $\theta_{23}$ on the old $\CPVA$ is symmetric with respect
to $\CPVA = 0$, while the new $\CPV$ has an odd-dependence on the old $\CPVA$.
%%%%%%%%%%%%%%%%%%%%%%%%%%%%

It is worth noting that the uncertainties of the mixing parameters in 
other representations cannot be evaluated through error propagation using 
the uncertainties of the mixing parameters obtained in the conventional 
representation. This is because the covariances of any two mixing parameters 
in the conventional representation are not provided. One way to obtain 
the uncertainties of the mixing parameters in any representation is through 
global fits of the experimental data using various representations. 
This will be reported in a separate article.
%%%%%%%%%%%%%%%%%%%%%%%%%%%%

% ============== Section{Transition Probability} =================== %
\section{Transition Probability} 
     \label{sec:NuParam_Prob}
     %\input{NuParam_41_Prob_v6.tex}

% -------- Contents ----------- %
Neutrinos and anti-neutrinos are related by a CP transformation which
transforms a left-handed neutrino into a right-handed anti-neutrino.
Recall Eq.(\ref{eq:NuParam_20}) that characterizes the neutrino oscillation
in the flavor eigenstate linking to the mass eigenstate through a unitary
mixing matrix. The coefficients of the massive anti-neutrino components are
simply related to the corresponding coefficients of the massive neutrino
components by complex conjugation. The anti-neutrinos can thus be described
by
\begin{equation}
\bar{\nu}_{\alpha}> = \sum_{k} U^{\star}_{\alpha k} |\bar{\nu}_{k}>  \; .
\label{eq:NuParam_40}
\end{equation}
The expressions for the transition probabilities of channels
$\alpha \rightarrow \beta$ in vacuum for neutrinos and anti-neutrinos
can be found in references, for example \cite{Giunti_Kim}, which are
listed as follows:
\begin{eqnarray}
& & P_{\nu_{\alpha} \rightarrow \nu_{\beta}}(L, E_{\nu})
    =  \delta_{\alpha \beta}                                  \nonumber \\ 
& & - 4 \sum_{k>j} Re\left[ U_{\alpha k} U^{\star}_{\beta k} 
                                 U^{\star}_{\alpha j} U_{\beta j} \right]
         \sin^2 \Delta_{kj}                                    \nonumber \\
& & + 2 \sum_{k>j} Im\left[ U_{\alpha k} U^{\star}_{\beta k} 
                             U^{\star}_{\alpha j} U_{\beta j} \right]
         \sin 2\Delta_{kj}                         \; , 
\label{eq:NuParam_prob_nu}
\end{eqnarray}
\begin{eqnarray}
& & P_{\bar{\nu}_{\alpha} \rightarrow \bar{\nu}_{\beta}}(L, E_{\nu})
    = \delta_{\alpha \beta}                                   \nonumber \\
& &  - 4 \sum_{k>j} Re\left[ U_{\alpha k} U^{\star}_{\beta k} 
                             U^{\star}_{\alpha j} U_{\beta j} \right]
        \sin^2 \Delta_{kj}                                     \nonumber \\
& & - 2 \sum_{k>j} Im\left[ U_{\alpha k} U^{\star}_{\beta k} 
                             U^{\star}_{\alpha j} U_{\beta j} \right]
       \sin 2\Delta_{kj}                          \; , 
\label{eq:NuParam_prob_antinu}
\end{eqnarray}
where
 $\Delta_{kj} = \Delta m^2_{kj} L/4 E_{\nu}  
            \equiv (m^2_k - m^2_j) L/4 E_{\nu}$
with $L$ and $E_{\nu}$ being the distance traveled and energy of neutrinos,
respectively.
The difference between these two transition probabilities appears only in
the sign of the imaginary parts that are quartic products of the elements
of the mixing matrix.
%%%%%%%%%%%%%%%%%%%%%%%%

Various neutrino oscillation experiments utilize different sources
of neutrinos and measure survival or transition probabilities.
Solar and reactor neutrino experiments observe survival probabilities of
$\nu_e$ or $\bar{\nu_e}$, whereas atmospheric and accelerator neutrino
experiments study $\nu_{\mu}$ disappearance,
$\nu_e$ appearance \cite{MINOS_numu_nue, T2K_numu_nue}, and $\nu_{\tau}$
appearance \cite{OPERA_numu_nutau}.  Given a neutrino oscillation channel,
some mixing parametrizations support simple probability forms, while
others give complicated expressions.
For the channel of $\nu_e \rightarrow \nu_e$, the survival probability
in representation $\RepA$ is
\begin{eqnarray}
& & \hskip-0.7cm
     \ProbA_{\nu_e \rightarrow \nu_e} =                          \nonumber \\
& & \hskip-0.7cm
     1 - (\cA_{13})^4 \sin^{2}2\thA_{12} \sin^{2}\Delta_{21}  
       - (\cA_{12})^2 \sin^{2}2\thA_{13} \sin^{2}\Delta_{31}     \nonumber \\
& & \hskip-0.7cm
       - (\sA_{12})^2 \sin^{2}2\thA_{13} \sin^{2}\Delta_{32}  \; ,
\label{eq:NuParam_prob_nue_A}   
\end{eqnarray}
while the survival probability in representation $\RepD$, for example, is
\begin{eqnarray}
& & \hskip-0.5cm
     \ProbD_{\nu_e \rightarrow \nu_e} = 1 -                     \nonumber \\
& & \hskip-0.5cm
   4 \left\{
      ( \cD_{23} \sD_{12} \cD_{13} - \sD_{23} \sD_{13} )^2
      ( \cD_{12} \cD_{13} )^2  \sin^{2}\Delta_{21}    \right.   \nonumber \\
& & \hskip-0.35cm
    + ( \sD_{23} \sD_{12} \cD_{13} + \cD_{23} \sD_{13} )^2 
      ( \cD_{23} \sD_{12} \cD_{13} - \sD_{23} \sD_{13} )^2 
                              \sin^{2}\Delta_{32}               \nonumber \\
& & \hskip-0.35cm
   \left. 
    + ( \sD_{23} \sD_{12} \cD_{13} + \cD_{23} \sD_{13} )^2
      ( \cD_{12} \cD_{13} )^2 \sin^{2}\Delta_{31}        
        \right\} ,
\label{eq:NuParam_prob_nue_D}   
\end{eqnarray}
which is considerably more complicated than
Eq.(\ref{eq:NuParam_prob_nue_A}).
Another example is the channel $\nu_{\mu} \rightarrow \nu_{\mu}$ studied
in the atmospheric or accelerator experiments. In representation
$\RepD$,
\begin{eqnarray}
& & \hskip-0.7cm
     \ProbD_{\nu_{\mu} \rightarrow \nu_{\mu}} =                  \nonumber \\
& & \hskip-0.7cm
     1 - (\cD_{23})^2 \sin^{2}2\thD_{12} \sin^{2}\Delta_{21}  
       - (\sD_{23})^2 \sin^{2}2\thD_{12} \sin^{2}\Delta_{31} \nonumber \\
& & \hskip-0.7cm
       - (\cD_{12})^4 \sin^{2}2\thD_{23} \sin^{2}\Delta_{32} , 
\label{eq:NuParam_prob_numu_D}
\end{eqnarray}
which is a simple expression, while in representation $\RepA$,
\begin{eqnarray}
& & \hskip-0.5cm
     \ProbA_{\nu_{\mu} \rightarrow \nu_{\mu}} = 1 -            \nonumber \\
& & \hskip-0.5cm
  4 \left\{
      ( \sA_{23} \sA_{12} \sA_{13} - \cA_{23} \cA_{12} )^2 
      ( \sA_{23} \cA_{12} \sA_{13} + \cA_{23} \sA_{12} )^2 
                 \sin^{2}\Delta_{21}  \right.                  \nonumber \\
& & \hskip-0.35cm
    + ( \sA_{23} \sA_{12} \sA_{13} - \cA_{23} \cA_{12} )^2
      ( \sA_{23} \cA_{13} )^2 \sin^{2}\Delta_{32}              \nonumber \\
& & \hskip-0.35cm
   \left. 
    + ( \sA_{23} \cA_{12} \sA_{13} + \cA_{23} \sA_{12} )^2
      ( \sA_{23} \cA_{13} )^2 \sin^{2}\Delta_{31}           
        \right\} ,
\label{eq:NuParam_prob_numu_A1}   
\end{eqnarray}
which is more complicated than that in representation $\RepD$. However,
in the limit of $\Delta_{31} \sim \Delta_{32} \gg \Delta_{21}$ and
$\thA_{13} \sim 0$, Eq.(\ref{eq:NuParam_prob_numu_A1}) can be reduced to
\begin{eqnarray}
 \ProbA_{\nu_{\mu} \rightarrow \nu_{\mu}} =  
       1 - \sin^2 2\thA_{23} \sin^{2}\Delta_{31} \; .
\label{eq:NuParam_prob_numu_A2}   
\end{eqnarray}
%%%%%%%%%%%%%%%%%%%%%%%%

In the limit of $\CPV = 0$,
Table~\ref{table:simpler_prob_form} lists representations in which the
survival or transition probabilities possess simpler forms in vacuum,
whose full expressions are given in Appendix~\ref{apdx:NuParam_Osci_Prob}.
\begin{table}[ht]
\begin{center}
\begin{tabular}{cc}  \hline\hline
    Probability                        &  Representations    \\ \hline
$P(\nu_e \rightarrow \nu_e)$           & $\RepA$, $\RepB$    \\
$P(\nu_{\mu} \rightarrow \nu_{\mu})$   & $\RepC$, $\RepD$    \\
$P(\nu_{\tau} \rightarrow \nu_{\tau})$ & $\RepE$, $\RepF$    \\
$P(\nu_{\mu} \rightarrow \nu_e)$       & $\RepA$, $\RepB$, $\RepC$, $\RepD$ \\
$P(\nu_{\mu} \rightarrow \nu_{\tau})$  & $\RepC$, $\RepD$, $\RepE$, $\RepF$ \\
                                                             \hline\hline
\end{tabular}
\caption{List of representations in which the survival or transition
         probabilities possess simpler forms in vacuum.}
\label{table:simpler_prob_form}
\end{center}
\end{table}
%
%
%%%%%%%%%%%%%%%%%%%%%%%%

% ============== Section{CP Asymmetry} ============================== %
\section{CP Asymmetry}
     \label{sec:NuParam_CP_Asym}
     %\input{NuParam_41_CP_Asym_v6.tex}

% -------- Contents ----------- %
For the case of $\CPV \ne 0$, the mixing matrix is complex and leads to
a violation of CP symmetry. Such a violation can be revealed by probing
the CP asymmetry, $A^{cp}_{\alpha \beta}$, in neutrino oscillation
experiments:
\begin{equation}
A^{cp}_{\alpha \beta} = 
     P_{\nu_{\alpha} \rightarrow \nu_{\beta}}  
   - P_{\bar{\nu}_{\alpha} \rightarrow \bar{\nu}_{\beta}}  \; .
\label{eq:NuParam_41}
\end{equation}
According to Eqs.(\ref{eq:NuParam_prob_nu}) and (\ref{eq:NuParam_prob_antinu}),
the CP asymmetry can be acquired readily:
\begin{equation}
A^{cp}_{\alpha \beta} = 
 4 \sum_{k>j} Im\left[ U_{\alpha k} U^{\star}_{\beta k} 
                         U^{\star}_{\alpha j} U_{\beta j} \right]
   \sin 2\Delta_{kj}    \; . 
\label{eq:NuParam_42_0}
\end{equation}
This expression confirms that the CP asymmetry can be probed only in the
transitions between different flavors since the imaginary part in
Eq.(\ref{eq:NuParam_42_0}) vanishes if $\alpha = \beta$.
For the oscillation channels $\nu_{\mu} \rightarrow \nu_e$ and
$\nu_{\mu} \rightarrow \nu_\tau$, the CP asymmetry in representation,
say $\RepS$, can be formulated as
\begin{equation}
A^{cp, \mathcal{S}}_{\alpha, \beta} = \kappa \cdot \frac{1}{2}
   \JpS \cdot \sin\CPVS \cdot \chi  \; ,
\label{eq:NuParam_42_1}
\end{equation}
where
\begin{equation*}
\JpS = 
   \sin 2\theta_{23}^{\mathcal{S}} \cdot \sin 2\theta_{12}^{\mathcal{S}} 
   \cdot \sin 2\theta_{13}^{\mathcal{S}} \cdot \cos\thS_{jk} \; ,
\end{equation*}
and
\begin{equation*}
\chi = \sin \left( \frac{\Delta m^2_{32} L}{2 E_{\nu}} \right)
     + \sin \left( \frac{\Delta m^2_{13} L}{2 E_{\nu}} \right)
     + \sin \left( \frac{\Delta m^2_{21} L}{2 E_{\nu}} \right) .
\end{equation*}
For $\nu_{\mu} \rightarrow \nu_e$,
\begin{eqnarray*}
 \kappa = 
   \left\{ 
      \begin{array}{ll}
        +1  & \mbox{for $\RepS$ = $\RepA$ or $\RepF$  }    \\
        -1  & \mbox{for $\RepS$ = $\RepB$, $\RepC$, $\RepD$, or $\RepE$}
      \end{array} 
   \right.,
\end{eqnarray*}
while for $\nu_{\mu} \rightarrow \nu_{\tau}$,
\begin{eqnarray*}
 \kappa = 
   \left\{ 
      \begin{array}{ll}
        -1  & \mbox{for $\RepS$ =  $\RepA$ or $\RepF$  }    \\
        +1  & \mbox{for $\RepS$ = $\RepB$, $\RepC$, $\RepD$, or $\RepE$}
      \end{array} 
   \right. .
\end{eqnarray*}
In Eq.(\ref{eq:NuParam_42_1}),
the $\theta_{jk}$ appearing in $\cos\theta_{jk}$ refers to the mixing
angle situated in the middle of the matrix product, $R_{ij} W_{jk} R_{ik}$.
%%%%%%%%%%%%%%%%%%%%%%%%%%%

% ------- Table{tbl_asym_th13_cp_v6.tex} ------ %
%\input{tbl_asym_th13_cp_v6.tex}
%
\begin{table}
\begin{minipage}[h]{1.0\linewidth}
\begin{tabular}{ccccccc} \hline\hline
$\thA_{13}$
    &  $\CPVA$  &  $\CPVB$  &   $\CPVC$   &  $\CPVD$   &  $\CPVE$ &  $\CPVF$
                                                          \\  \hline
 $1.0$
   &   5.0  &  -5.001  &  -4.933  &  0.319   &  0.251   & -0.317  \\
   &  50.0  & -50.008  & -49.398  &  2.798   &  2.189   & -2.790  \\
   & -35.0  &  35.007  &  34.554  & -2.096   & -1.643   &  2.089  \\ \hline
 $5.44$
   &   5.0  &  -5.041  &  -4.722  &   1.778  &   1.460  &   -1.737  \\
   &  50.0  & -50.229  & -47.256  &  15.358  &  12.384  &  -15.128  \\
   & -35.0  &  35.219  &  33.060  & -11.590  &  -9.431  &   11.371  \\ \hline
 $9.44$
   &   5.0  &  -5.123  &  -4.638  &   3.209  &   2.725  &   -3.087  \\
   &  50.0  & -50.685  & -46.015  &  26.882  &  22.212  &  -26.197  \\
   & -35.0  &  35.658  &  32.324  & -20.559  & -17.225  &   19.902  \\ \hline
 $15.0$
   &   5.0  &  -5.310  &  -4.661  &   5.556  &   4.907  &   -5.246  \\
   &  50.0  & -51.699  &  -45.232 &  43.239  &  36.772  &  -41.540  \\
   & -35.0  &  36.646  &   32.108 & -34.047  & -29.508  &   32.401  \\
                                                                \hline \hline
\end{tabular}
\caption{Expected $\CPV$ values in various representatsions
         for given $(\thA_{13}, \CPVA)$, where $\thA_{12} = 33.91$
         and $\thA_{23} = 45$ are used. Note that all numbers in this
         table are in degrees.}
\label{table:Asym_th13_CPphase} \vspace{-0.4cm}
\end{minipage}
\end{table}
The CP asymmetry is a physical observable and has been verified to be
invariant in all representations.
Given $(\thA_{13}, \CPVA)$ in the conventional representation $\RepA$,
Table~\ref{table:Asym_th13_CPphase} shows the expected $\CPV$ values
in other representations.
From Table~\ref{table:NuParam_zero_CP_ex}, it can be seen
that the three mixing angles in representations $\RepD$, $\RepE$, $\RepF$
are far away from zero unlike those in representations $\RepA$, $\RepB$,
$\RepC$ in which $\theta_{13}$ is small. Given
that the CP asymmetry is invariant and the values of $\Jp$ are larger in
representations $\RepD$, $\RepE$, $\RepF$ than those in representations
$\RepA$, $\RepB$, $\RepC$, the $\CPV$ values in representations
$\RepD$, $\RepE$, $\RepF$, would be smaller than those in
representations $\RepA$, $\RepB$, $\RepC$.
However, if $\theta_{13}$ should turn out to be so small that
the upper limit can only be obtained from experiments,
this would not allow the determination of
$\CPV$ in the conventional representation because
there are two unknown parameters (i.e., $\theta_{13}$ and $\CPV$) in the
CP-violation observable [see Eq.(\ref{eq:NuParam_42_1})].
On the other hand, representations $\RepD$, $\RepE$, $\RepF$ produce
large values of $\theta_{ij}$ and thus there is only one unknown parameter
(i.e., $\CPV$), which makes it possible
to determine $\CPV$.
%%%%%%%%%%%%%%%%%%%%%%%%%%%%

% ==================== Section{Conclusions} ======================== %
\section{Conclusions}
     \label{sec:NuParam_concl}
     %\input{NuParam_71_concl_v6.tex}

% -------- Contents ---------- %
We have studied several different parametrizations for the neutrino
mixing matrix corresponding to different mixing angles and CP-violating
phases.  For both cases of $\CPV = 0$ and $\CPV \ne 0$, the transformation 
of $(\theta_{ij}, \CPV)$ between two representations is derived.
For the $\CPV = 0$ case, we present the predicted $\theta_{ij}$ values 
in various representations.
For the $\CPV \ne 0$ case, we show how $\CPV$ can impact on the transformation 
of $(\theta_{ij}, \CPV)$ from one representation to another.
Solving for $\theta_{ij}$ in the various mixing parameterizations has
shown that representations $\RepD$, $\RepE$, and $\RepF$ produce significant
non-zero $\theta_{ij}$. This suggests that these three representations
are more suitable for probing $\CPV$.
We have also examined how the survival and
transition probabilities depend on the mixing matrix representation,
and identified the representations and oscillation channels for which
simpler expressions exist.
%%%%%%%%%%%%%%%%%%%%%%

In conclusion, the mixing matrix describing the neutrino oscillation is
unique, but the structure of each element of the mixing matrix vary,
depending on the parametrizations. That is, the $(\theta_{ij}, \CPV)$
values vary from one representation to another. In the conventional
representation, $\theta_{13}$ is believed to be small or zero. This work
reports alternative parametrizations for the mixing matrix that
can produce significant non-zero mixing parameters and thus provides
an easier way for probing $\CPV$.
%%%%%%%%%%%%%%%%%%%%%% 

% ================ Section{Acknowledgements} ======================= %
\section{Acknowledgements}
     %\input{NuParam_81_thanks_v6.tex}

% ---------- Contents ----------- %
This research was supported by Taiwan National Science Council under Project
No. NSC 98-2811-M-002-501 and NSC 99-2811-M-002-064, Canadian Natural Sciences
and Engineering Research Council, U.S. National Science Foundation, and
Fermi Research Alliance, LLC under the U.S. Department of Energy contract
No. DE-AC02-07CH11359. One of the authors (J.P.) thanks the members of the
Insitute of Physics, Academia Sinica for their hospitality.

% ==================== Begin Appendix ============================== %
\appendix

% ======= Section{Solutions of $\theta_{ij}$ for $\CPV = 0$ Case } ====== % 
\section{Solutions of $\theta_{ij}$ for $\CPV = 0$ Case }
     \label{apdx:NuParam_UMatrix_zeroCP}
     %\input{z_apdx_NuParam_UMatrix_zeroCP_v6.tex}

% -------- Contents --------- %
Below presents the mixng matrix, $U$, in different representations for the
case of $\CPV = 0$. In the conventional representation $\RepA$,
\begin{eqnarray*}
& & \hskip-0.3cm
  U = R_{x}(\thA_{23}) R_{y}(\thA_{13}) R_{z}(\thA_{12}) 
    =                                            \nonumber \\
& &  \hskip-0.3cm
  \left(
    \begin{array}{ccc}
      \cA_{12}\cA_{13}   &       \sA_{12}\cA_{13}    &  \sA_{13}     \\
     -\cA_{23}\sA_{12}-\sA_{23}\cA_{12}\sA_{13}  &
      \cA_{23}\cA_{12}-\sA_{23}\sA_{12}\sA_{13}  & \sA_{23}\cA_{13}  \\
      \sA_{23}\sA_{12}-\cA_{23}\cA_{12}\sA_{13}  &
     -\sA_{23}\cA_{12}-\cA_{23}\sA_{12}\sA_{13}  &  \cA_{23}\cA_{13} 
    \end{array}
  \right) .
\end{eqnarray*}
%
%%%%%%%%%%%%%%%%%%%%%
%
In the representation $\RepB$,
\begin{eqnarray*}
& & \hskip-0.3cm
  U = R_{x}(\thB_{23}) R_{z}(\thB_{12}) R_{y}(\thB_{13}) 
    =                                            \nonumber \\
& & \hskip-0.3cm
  \left(
    \begin{array}{ccc}
      \cB_{12}\cB_{13}   &       \sB_{12}        &  \cB_{12}\sB_{13}     \\
     -\sB_{23}\sB_{13}-\cB_{23}\sB_{12}\cB_{13}  &  \cB_{23}\cB_{12}  &  
      \sB_{23}\cB_{13}-\cB_{23}\sB_{12}\sB_{13}    \\
     -\cB_{23}\sB_{13}+\sB_{23}\sB_{12}\cB_{13}  & -\sB_{23}\cB_{12}  & 
      \cB_{23}\cB_{13}+\sB_{23}\sB_{12}\sB_{13} 
    \end{array}
  \right) .
\end{eqnarray*}
%
%%%%%%%%%%%%%%%%%%%%%
%
In the representation $\RepC$,
\begin{eqnarray*}
& & \hskip-0.3cm
  U = R_{y}(\thC_{13}) R_{x}(\thC_{23}) R_{z}(\thC_{12}) 
    =                                            \nonumber \\
& & \hskip-0.3cm
  \left(
    \begin{array}{ccc}
      \cC_{12}\cC_{13}+\sC_{23}\sC_{12}\sC_{13}  &    
      \sC_{12}\cC_{13}-\sC_{23}\cC_{12}\sC_{13}  &  \cC_{23}\sC_{13}    \\  
     -\cC_{23}\sC_{12}   &  \cC_{23}\cC_{12}     &  \sC_{23}  \\
     -\cC_{12}\sC_{13}+\sC_{23}\sC_{12}\cC_{13}  &    
     -\sC_{12}\sC_{13}-\sC_{23}\cC_{12}\cC_{13}  &  \cC_{23}\cC_{13}   
    \end{array}
  \right) .
\end{eqnarray*}
%
%%%%%%%%%%%%%%%%%%%%%%%
%
In the representation $\RepD$,
\begin{eqnarray*}
& & \hskip-0.3cm
  U = R_{y}(\thD_{13}) R_{z}(\thD_{12}) R_{x}(\thD_{23}) 
    =                                            \nonumber \\
& & \hskip-0.3cm
  \left(
    \begin{array}{ccc}
      \cD_{12}\cD_{13}   &  -\sD_{23}\sD_{13}+\cD_{23}\sD_{12}\cD_{13}  &
      \cD_{23}\sD_{13}+\sD_{23}\sD_{12}\cD_{13}            \\
     -\sD_{12}    &   \cD_{23}\cD_{12}    &  \sD_{23}\cD_{12}   \\
     -\cD_{12}\sD_{13}   &  -\sD_{23}\cD_{13}-\cD_{23}\sD_{12}\sD_{13}  &
      \cD_{23}\cD_{13}-\sD_{23}\sD_{12}\sD_{13}     
    \end{array}
  \right) .
\end{eqnarray*}
%
%%%%%%%%%%%%%%%%%%%%%
%
In the representation $\RepE$,
\begin{eqnarray*}
& & \hskip-0.3cm
  U = R_{z}(\thE_{12}) R_{x}(\thE_{23}) R_{y}(\thE_{13}) 
    =                                            \nonumber \\
& & \hskip-0.3cm
  \left(
    \begin{array}{ccc}
      \cE_{12}\cE_{13}-\sE_{23}\sE_{12}\sE_{13}  &   \cE_{23}\sE_{12}   &
      \cE_{12}\sE_{13}+\sE_{23}\sE_{12}\cE_{13}            \\
     -\sE_{12}\cE_{13}-\sE_{23}\cE_{12}\sE_{13}  &   \cE_{23}\cE_{12}   &
     -\sE_{12}\sE_{13}+\sE_{23}\cE_{12}\cE_{13}            \\
     -\cE_{23}\sE_{13}  &  -\sE_{23}   &   \cE_{23}\cE_{13}
    \end{array}
  \right) .
\end{eqnarray*}
%
%%%%%%%%%%%%%%%%%%%%
%
In the representation $\RepF$,
\begin{eqnarray*}
& & \hskip-0.3cm
  U = R_{z}(\thF_{12}) R_{y}(\thF_{13}) R_{x}(\thF_{23}) 
    =                                            \nonumber \\
& & \hskip-0.3cm
  \left(
    \begin{array}{ccc}
      \cF_{12}\cF_{13}   &  \cF_{23}\sF_{12}-\sF_{23}\cF_{12}\sF_{13}   &
      \sF_{23}\sF_{12}+\cF_{23}\cF_{12}\sF_{13}           \\
     -\sF_{12}\cF_{13}   &  \cF_{23}\cF_{12}+\sF_{23}\sF_{12}\sF_{13}   &
      \sF_{23}\cF_{12}-\cF_{23}\sF_{12}\sF_{13}           \\
     -\sF_{13}  &  -\sF_{23}\cF_{13}   &   \cF_{23}\cF_{13}
    \end{array}
  \right) .
\end{eqnarray*}
%
%%%%%%%%%%%%%%%%%%
%
Denote the unitary matrix as
\begin{eqnarray*}
 U &=&  
  \left(
    \begin{array}{ccc}
    U_{e1}       &   U_{e2}        &   U_{e3}       \\
    U_{\mu 1}    &   U_{\mu 2}     &   U_{\mu 3}     \\
    U_{\tau 1}   &   U_{\tau 2}    &   U_{\tau 3}    \\
    \end{array}
  \right)  .
\end{eqnarray*}
Through the elements of the unitary matrix, one can solve the three mixing
angles in each representation, as shown in
Table~\ref{table:z_apdx_NuParam_sol_zeroCP}.
\begin{table}[ht]
\begin{center}
\begin{tabular}{c|ll} \hline\hline
Rep.  &   \multicolumn{2}{c}{Solutions}   \\ \hline
$\RepA$
     &  $\sA_{13} = U_{e3} \Rightarrow \thA_{13}$ ;  &
        $\sA_{23}\cA_{13} = U_{\mu 3} \Rightarrow \thA_{23}$ ; \\
  &  &  $\sA_{12}\cA_{13} = U_{e2} \Rightarrow \thA_{12}$      \\
$\RepB$
     &  $\sB_{12} = U_{e2} \Rightarrow \thB_{12}$ ;   &
        $\sB_{23}\cB_{12} = -U_{\tau 2} \Rightarrow \thB_{23}$ ;  \\
  &  &  $\cB_{12}\sB_{13} = U_{e3} \Rightarrow \thB_{13}$         \\
$\RepC$
     &  $\sC_{23} = U_{\mu 3} \Rightarrow \thC_{23}$ ;  &
        $\cC_{23}\sC_{13} = U_{e3} \Rightarrow \thC_{13}$ ;       \\
  &  &  $\cC_{23}\sC_{12} = -U_{\mu 1} \Rightarrow \thC_{12}$     \\
$\RepD$
     &  $\sD_{12} = -U_{\mu 1} \Rightarrow \thD_{12}$ ;  &
        $\sD_{23}\cD_{12} = U_{\mu 3} \Rightarrow \thD_{23}$ ;    \\
  &  &  $\cD_{12}\sD_{13} = -U_{\tau 1} \Rightarrow \thD_{13}$    \\
$\RepE$
     &  $\sE_{23} = -U_{\tau 2} \Rightarrow \thE_{23}$ ;  &
        $\cE_{23}\sE_{12} = U_{e2} \Rightarrow \thE_{12}$ ;       \\
  &  &  $\cE_{23}\sE_{13} = -U_{\tau 1} \Rightarrow \thE_{13}$    \\
$\RepF$
     &  $\sF_{13} = -U_{\tau 1} \Rightarrow \thF_{13}$ ;  &
        $\sF_{23}\cF_{13} = -U_{\tau 2} \Rightarrow \thF_{23}$ ;  \\
  &  &  $\sF_{12}\cF_{13} = -U_{\mu 1} \Rightarrow \thF_{12}$
                                                         \\ \hline \hline
\end{tabular}
\caption{Solutions of the three mixing angles in different representatsions
         for the case of $\CPV = 0$.}
\label{table:z_apdx_NuParam_sol_zeroCP} \vspace{-0.4cm}
\end{center}
\end{table}
%%%%%%%%%%%%%%%%%%%%%%%%

% === Section{Solutions of $(\theta_{ij}, \CPV)$ for $\CPV \ne 0$ Case} == %
\section{Solutions of $(\theta_{ij}, \CPV)$ for $\CPV \ne 0$ Case }  
     \label{apdx:NuParam_Solution_0}
     %\input{z_apdx_NuParam_Sol_20_v6.tex}

% ------- Contents -------- %
The procedure of obtaining $(\theta_{ij}, \CPV)$ for each representation
as expressed in terms of $(\thA_{ij}, \CPVA)$ is described in details in
Section~\ref{apdx:NuParam_Solution_AD} using transformation from
representation $\RepA$ to $\RepD$. Following the procedure presented
in Section~\ref{apdx:NuParam_Solution_AD}, the transformation from
representation $\RepA$ to other representations is briefly summarized
in this appendix by only listing the equations of the nine parameters.
%%%%%%%%%%%%%%%%%%%%%%%%%

% =========== SubSection{$(\thD_{ij}, \CPVD)$ Solutions } =========== %
\subsection{$(\thD_{ij}, \CPVD)$ Solutions }
      \label{apdx:NuParam_Solution_AD}
      %\input{z_apdx_NuParam_Sol_2_AD_v6.tex}

% -------- Contents ---------- %
Rotations undertaken with the matrices presented in Eq.(\ref{eq:NonZeroCP_31})
will have nine independent parameters:  five phases in $D^L$ and $D^R$, three
mixing angles and one $\CPV$ phase from the mixing matrix.  To solve for
$(\thD_{ij}, \CPVD)$ in representation $\RepD$ where the
$(\thA_{ij}, \CPVA)$ are known for representation $\RepA$, start with
Eq.(\ref{eq:NuParam_39_1}):
%%%
\begin{equation}
U = RWR(\thD_{ij}, \CPVD) = D^L \cdot RWR(\thA_{ij}, \CPVA) \cdot D^R \; ,
\end{equation}
\label{eq:Append_D_AD_1}
%%%
where
\begin{eqnarray}
D^L(\PhiD_{Li}) 
    &=& diag \left( e^{i\PhiD_{L1}}, e^{i\PhiD_{L2}}, e^{i\PhiD_{L3}} \right)  
                                                     \; ,  \nonumber \\
D^R(\Phi_{Ri})
    &=& diag \left( e^{i\PhiD_{R1}}, e^{i\PhiD_{R2}}, 1  \right)   \; .  
\label{eq:Append_D_AD_2}
\end{eqnarray}
%%%
The nine real parts and the nine imaginary parts of
Equation~(\ref{eq:Append_D_AD_1}) are listed in
Eqs.(\ref{eq:Append_D_AD_31}) through (\ref{eq:Append_D_AD_39}).
For element (1,1):
\begin{eqnarray}
\cD_{13} \cD_{12}  &=&  \cA_{13} \cA_{12} \cos(\PhiD_{L1} + \PhiD_{R1}) , 
                                                     \nonumber \\
0  &=&  \cA_{13} \cA_{12} \sin(\PhiD_{L1} + \PhiD_{R1}) . 
\label{eq:Append_D_AD_31}
\end{eqnarray}
For element (2,1):
\begin{eqnarray}
& & \hskip-1.5cm
\sD_{12} \cos\CPVD = \cA_{23} \sA_{12} \cos(\PhiD_{L2} + \PhiD_{R1})  
                                                     \nonumber \\
& & \hskip0.5cm 
     + \sA_{23} \cA_{12} \sA_{13} \cos(\CPVA + \PhiD_{L2} + \PhiD_{R1}) ,
                                                     \nonumber \\
& & \hskip-1.5cm
\sD_{12} \sin\CPVD = \cA_{23} \sA_{12} \sin(\PhiD_{L2} + \PhiD_{R1})  
                                                     \nonumber \\
& & \hskip0.5cm
     + \sA_{23} \cA_{12} \sA_{13} \sin(\CPVA + \PhiD_{L2} + \PhiD_{R1}) .
\label{eq:Append_D_AD_32}
\end{eqnarray}
For element (3,1):
\begin{eqnarray}
& & \hskip-1.5cm
  -\sD_{13} \cD_{12} = \sA_{23} \sA_{12} \cos(\PhiD_{L3} + \PhiD_{R1}) 
                                                     \nonumber \\ 
& & \hskip0.5cm
     -\cA_{23} \cA_{12} \sA_{13} \cos(\CPVA + \PhiD_{L3} + \PhiD_{R1}) ,
                                                     \nonumber \\
& & \hskip-0.5cm
  0  =  \sA_{23} \sA_{12} \sin(\PhiD_{L3} + \PhiD_{R1})  \nonumber \\
& & \hskip0.5cm
    -\cA_{23} \cA_{12} \sA_{13} \sin(\CPVA + \PhiD_{L3} + \PhiD_{R1}) .
\label{eq:Append_D_AD_33}
\end{eqnarray}
For element (1,2):
\begin{eqnarray}
\hskip-0.7cm
\cD_{23} \cD_{13} \sD_{12} \cos\CPVD -\sD_{23} \sD_{13}
                  &=&  \sA_{12} \cA_{13} \cos(\PhiD_{L1} + \PhiD_{R2}) , 
                                                     \nonumber \\
-\cD_{23} \cD_{13} \sD_{12} \sin\CPVD
                  &=&  \sA_{12} \cA_{13} \sin(\PhiD_{L1} + \PhiD_{R2}) . 
\label{eq:Append_D_AD_34}
\end{eqnarray}
For element (2,2):
\begin{eqnarray}
& & \hskip-1.5cm
  \cD_{23} \cD_{12} = \cA_{23} \cA_{12} \cos(\PhiD_{L2} + \PhiD_{R2})
                                                        \nonumber \\
& & \hskip0.5cm
    -\sA_{23} \sA_{12} \sA_{13} \cos(\CPVA + \PhiD_{L2} + \PhiD_{R2}) , 
                                                        \nonumber \\
& & \hskip-0.5cm
  0 = \cA_{23} \cA_{12} \sin(\PhiD_{L2} + \PhiD_{R2})   \nonumber \\
& & \hskip0.5cm
    -\sA_{23} \sA_{12} \sA_{13} \sin(\CPVA + \PhiD_{L2} + \PhiD_{R2}) .
\label{eq:Append_D_AD_35}
\end{eqnarray}
For element (3,2):
\begin{eqnarray}
& & \hskip-1.5cm
 \sD_{23} \cD_{13} + \cD_{23} \sD_{12} \sD_{13} \cos\CPVD
      = \sA_{23} \cA_{12} \cos(\PhiD_{L3} + \PhiD_{R2})  \nonumber \\
& & \hskip0.7cm  
      +\cA_{23} \sA_{12} \sA_{13} \cos(\CPVA + \PhiD_{L3} + \PhiD_{R2}) ,
                                                             \nonumber \\
& & \hskip-1.5cm
 -\cD_{23} \sD_{12} \sD_{13} \sin\CPVD
      = \sA_{23} \cA_{12} \sin(\PhiD_{L3} + \PhiD_{R2})   \nonumber \\
& & \hskip0.7cm
      +\cA_{23} \sA_{12} \sA_{13} \cos(\CPVA + \PhiD_{L3} + \PhiD_{R2}) . 
\label{eq:Append_D_AD_36}
\end{eqnarray}
For element (1,3):
\begin{eqnarray}
& & \hskip-1.5cm
  \cD_{23} \sD_{13} + \sD_{23} \sD_{12} \cD_{13} \cos\CPVD
                 = \sA_{13} \cos(\CPVA - \PhiD_{L1} ) ,
                                                     \nonumber \\
& & \hskip-1.5cm
 -\sD_{23} \sD_{12} \cD_{13} \sin\CPVD
                 = -\sA_{13} \sin(\CPVA - \PhiD_{L1} ) .
\label{eq:Append_D_AD_37}
\end{eqnarray}
For element (2,3):
\begin{eqnarray}
\sD_{23} \cD_{12} &=& \sA_{23} \cA_{13} \cos\PhiD_{L2} ,   \nonumber \\
  0  &=& \sA_{23} \cA_{13} \sin\PhiD_{L2} . 
\label{eq:Append_D_AD_38}
\end{eqnarray}
For element (3,3):
\begin{eqnarray}
& & \hskip-1.5cm
  \cD_{23} \cD_{13} - \sD_{23} \sD_{12} \sD_{13} \cos\CPVD
                 = \cA_{23} \cA_{13} \cos\PhiD_{L3} ,   \nonumber \\
& & \hskip-1.5cm
  \sD_{23} \sD_{12} \sD_{13} \sin\CPVD
                 = \cA_{23} \cA_{13} \sin\PhiD_{L3} . 
\label{eq:Append_D_AD_39}
\end{eqnarray}
%%%%%%%%%%%%%%%%%%%%%%%%%

Based on Eqs.(\ref{eq:Append_D_AD_31})--(\ref{eq:Append_D_AD_39}), one
can solve $(\thD_{ij}, \CPVD)$ in terms of $(\thA_{ij}, \CPVA)$ as detailed
in the following.
From the imaginary parts of elements (1,1) and (2,3), one has
\begin{eqnarray}
\PhiD_{L1} + \PhiD_{R1} &=& 0 ,   \nonumber \\
\PhiD_{L2} &=&  0 .
\label{eq:Append_D_AD_41}
\end{eqnarray}
%%%%%%%
From the imaginary part of element (2,2), one has
\begin{equation}
\sin^2{\PhiD_{R2}} = \frac{ (a~\sin\CPVA )^2}
                          { (a~\sin\CPVA )^2 + (a~\cos\CPVA - b)^2}  \; ,
\label{eq:Append_D_AD_42}
\end{equation}
where $a = \sA_{23} \sA_{12} \sA_{13}$ and $b = \cA_{23} \cA_{12}$ .
The sign of $\PhiD_{R2}$ can be easily determined by the original
equation, Eq.(\ref{eq:Append_D_AD_35}).
%%%%%%%
From the imaginary part of element (3,1), one has
\begin{equation}
\sin^2{\PhiD_{34}} = 
      \frac{ (a^{'} \sin\CPVA )^2 }
           { (a^{'} \sin\CPVA )^2 + (a^{'} \cos\CPVA - b^{'})^2 } \; ,
\label{eq:Append_D_AD_43}
\end{equation}
where $a^{'} = \cA_{23} \cA_{12} \sA_{13}$,  $b^{'} = \sA_{23} \sA_{12}$,
and $\PhiD_{34} \equiv \PhiD_{L3} + \PhiD_{R1}$. Again, one can readily
determine the sign of $\PhiD_{34}$ using the original equation,
Eq.(\ref{eq:Append_D_AD_33}), that is used for obtaining
Eq.(\ref{eq:Append_D_AD_43}).
%%%%%%%
With these phases, the three mixing angles in representation $\RepD$ can be
extracted using the real parts of elements (1,1), (3,1), (2,2), and (2,3):
\begin{eqnarray}
& & \hskip-1.2cm
  \tan \thD_{23} =
       \frac{\sA_{23} \cA_{13}}
            {\cA_{23} \cA_{12} \cos\PhiD_{R2} 
                - \sA_{23} \sA_{12}\sA_{13}\cos(\CPVA + \PhiD_{R2})} ,
\label{eq:Append_D_AD_44}
\end{eqnarray}
\begin{eqnarray}
& & \hskip-1.2cm
  \cos \thD_{12} = 
       \frac{\sA_{23} \cA_{13}}{\sD_{23}}  \nonumber \\
& & \hskip-0.3cm
   = \frac{\cA_{23}\cA_{12}\cos\PhiD_{R2} 
                - \sA_{23}\sA_{12}\sA_{13}\cos(\CPVA + \PhiD_{R2})}
          {\cD_{23}} , 
\label{eq:Append_D_AD_45}
\end{eqnarray}
\begin{eqnarray}
& & \hskip-1.2cm
  \sin \thD_{13} = 
      \frac{ \cA_{23}\cA_{12}\sA_{13}\cos(\CPVA + \PhiD_{34})
           - \sA_{23}\sA_{12}\cos\PhiD_{34}} {\cD_{12}}.
\label{eq:Append_D_AD_46}
\end{eqnarray}
%%%%%%%
The remaining parameters thus can be determined.
From the real and imaginary parts of element (1,3), one has
\begin{eqnarray}
& & \hskip-1.2cm
  \cos(\CPVA - \PhiD_{L1}) = 
       \frac{( \cD_{23}\sD_{13} )^2 + ( \sA_{13} )^2 
                                    - ( \sD_{23}\sD_{12}\cD_{13} )^2 }  
            {2 \cD_{23}\sD_{13} \sA_{13} }.
\label{eq:Append_D_AD_47}
\end{eqnarray}
As before, the sign of $\CPVA - \PhiD_{L1}$ can be determined by the real
and imaginary parts of Eq.(\ref{eq:Append_D_AD_33}) or
(\ref{eq:Append_D_AD_34}), and thus $\PhiD_{L1}$ can be determined.
%%%%%%%
Subsequently,
$\PhiD_{R1} = - \PhiD_{L1}$ and $\PhiD_{L3} = \PhiD_{34} - \PhiD_{R1}$ can
be therefore determined. Finally, the CP-violating phase in representation
$\RepD$, $\CPVD$, can be resolved using those conditions associated with
$\sin \CPVD$ in elements (2,1), (1,2), (3,2), (1,3), or (3,3), which are
listed as follows:
\begin{eqnarray}
& & \hskip-1.0cm 
  \sin \CPVD                                \nonumber \\
& & \hskip-1.0cm
  = \frac{ \sA_{23}\cA_{12}\sA_{13} \sin(\CPVA + \eta) 
          +\cA_{23}\sA_{12} \sin \eta } { \sD_{12} } 
                                            \nonumber \\ 
& & \hskip-1.0cm
  = \frac{ \cA_{23}\sA_{12}\sA_{13} \sin(\CPVA + \xi) 
          +\sA_{23}\cA_{12} \sin \xi } {-\cD_{23}\sD_{12}\sD_{13} } 
                                            \nonumber \\ 
& & \hskip-1.0cm
  = \frac{-\sA_{12}\cA_{13} \sin(\PhiD_{L1} + \PhiD_{R2}) }  
         { \cD_{23}\sD_{12}\cD_{13} }        \nonumber \\
& & \hskip-1.0cm
  = \frac{\sA_{13} \sin(\CPVA - \PhiD_{L1}) }  
         {\sD_{23}\sD_{12}\cD_{13} }        \nonumber \\
& & \hskip-1.0cm
  = \frac{\cA_{23}\cA_{13} \sin \PhiD_{L3} }  
         {\sD_{23}\sD_{12}\sD_{13} }  \; ,  
\label{eq:Append_D_AD_5}
\end{eqnarray}
where $\eta \equiv \PhiD_{L2} + \PhiD_{R1}$  and
$\xi  \equiv \PhiD_{L3} + \PhiD_{R2}$.
Consistency in the value of $\sin \CPVD$ calculated from the five different
expressions in Eq.(\ref{eq:Append_D_AD_5}) serves as a means to check whether
the values of the nine parameters are correct.  Furthermore,
by looking at Eq.(\ref{eq:Append_D_AD_1}), 18 conditions are formed, nine
from real parts and the other nine from imaginary parts.
With the solutions presented in Eqs.(\ref{eq:Append_D_AD_41}) through
(\ref{eq:Append_D_AD_5}), the consistency among the 18 conditions has
been checked.
%%%%%%%%%%%%%%%%%%%%%%%%%%%%%%

% ========== SubSection{$(\thB_{ij}, \CPVB)$ Solutions } ============= %
\subsection{$(\thB_{ij}, \CPVB)$ Solutions }
        \label{apdx:NuParam_Solution_AB}
        %\input{z_apdx_NuParam_Sol_2_AB_v6.tex}

% -------- Contents ----------- %
For the rotation matrices in Eq.(\ref{eq:NonZeroCP_31}) to transform 
$(\thA_{ij}, \CPVA)$ from representations $\RepA$ to $\RepB$, the solutions 
to the nine parameters are listed as follows.
\begin{equation}
\PhiB_{R1} = -\CPVA  \hskip0.5cm \rm{and}  \hskip0.5cm
\PhiB_{L1} =  \CPVA \; .
\label{eq:Append_C_AB_1}
\end{equation}
\begin{equation}
\sin^2{\PhiB_{25}} = \frac{ (a~\sin\CPVA )^2}
                           { (a~\sin\CPVA )^2 + (a~\cos\CPVA - b)^2}  \; ,
\label{eq:Append_C_AB_2}
\end{equation}
where $a = \sA_{23} \sA_{12} \sA_{13}$, $b = \cA_{23} \cA_{12}$, and
$\PhiB_{25} = \PhiB_{L2} + \PhiB_{R2}$ .
\begin{equation}
\sin^2{\PhiB_{35}} = 
      \frac{ (a^{'} \sin\CPVA )^2 }
           { (a^{'} \sin\CPVA )^2 + (a^{'} \cos\CPVA + b^{'})^2 } \; ,
\label{eq:Append_C_AB_3}
\end{equation} 
where $a^{'} = \cA_{23} \sA_{12} \sA_{13}$,  $b^{'} = \sA_{23} \cA_{12}$, 
and $\PhiB_{35} \equiv \PhiB_{L3} + \PhiB_{R2}$ . 
The three mixing angles can thus be obtained in representation $\RepB$. 
\begin{eqnarray}
& & \hskip-1.3cm
  \tan \thB_{23} =
       \frac{\sA_{23}\cA_{12}\cos\PhiB_{35} 
                + \cA_{23}\sA_{12}\sA_{13}\cos(\CPVA + \PhiB_{35})}
            {\cA_{23}\cA_{12}\cos\PhiB_{25} 
                - \sA_{23}\sA_{12}\sA_{13}\cos(\CPVA + \PhiB_{25})} ,
\label{eq:Append_C_AB_4}
\end{eqnarray}
\begin{eqnarray}   
& & \hskip-1.3cm
  \cos \thB_{12} = 
       \frac{\sA_{23}\cA_{12}\cos\PhiB_{35} 
                + \cA_{23}\sA_{12}\sA_{13}\cos(\CPVA + \PhiB_{35})}
            {\sB_{23}}                                  \nonumber \\
& & \hskip-0.2cm
     = \frac{\cA_{23}\cA_{12}\cos\PhiB_{25} 
                - \sA_{23}\sA_{12}\sA_{13}\cos(\CPVA + \PhiB_{25})}
            {\cB_{23}} ,
\label{eq:Append_C_AB_5}
\end{eqnarray}
\begin{eqnarray}
& & 
  \sin \thB_{13} = 
       \frac{\sA_{13}} {\cB_{12}} \; .
\label{eq:Append_C_AB_6}
\end{eqnarray}
The remaining parameters thus can be determined as follows.
\begin{eqnarray}
& & \hskip-1.2cm
  \cos \PhiB_{L2} = 
       \frac{( \sB_{23}\cB_{13} )^2 + ( \sA_{23} \cA_{13} )^2 
                                    - ( \cB_{23}\sB_{12}\sB_{13} )^2 }  
            {2 \sB_{23}\cB_{13} \sA_{23}\cA_{13} }  \; .
\label{eq:Append_C_AB_7}
\end{eqnarray}
Therefore, 
$\PhiB_{R2} = \PhiB_{25} - \PhiB_{L2}$ and 
$\PhiB_{L3} = \PhiB_{35} - \PhiB_{R2}$ can be determined. 
Finally, the CP-violating phase in representation $\RepB$, $\CPVB$, can be 
obtained using those conditions associated with $\sin \CPVB$ :
\begin{eqnarray}
& & \hskip-1.0cm 
  \sin \CPVB                                \nonumber \\
& & \hskip-1.0cm
  = \frac{ \sA_{23}\cA_{12}\sA_{13} \sin(\CPVA + \etaB)  
          +\cA_{23}\sA_{12} \sin \etaB } { \cB_{23}\sB_{12}\cB_{13} } 
                                            \nonumber \\ 
& & \hskip-1.0cm
  = \frac{-\cA_{23}\cA_{12}\sA_{13} \sin(\CPVA + \xiB) 
          +\sA_{23}\sA_{12} \sin \xiB } { \sB_{23}\sB_{12}\cB_{13} } 
                                            \nonumber \\ 
& & \hskip-1.0cm
  = \frac{-\sA_{12}\cA_{13} \sin(\PhiB_{L1} + \PhiB_{R2}) }  
         {\sB_{12} }                        \nonumber \\
& & \hskip-1.0cm
  = \frac{-\sA_{23}\cA_{13} \sin \PhiB_{L2} }  
         {\cB_{23}\sB_{12}\sB_{13} }        \nonumber \\
& & \hskip-1.0cm
  = \frac{\cA_{23}\cA_{13} \sin \PhiB_{L3} }  
         {\sB_{23}\sB_{12}\sB_{13} }  \; ,  
\label{eq:Append_C_AB_8}
\end{eqnarray}
where $\etaB \equiv \PhiB_{L2} + \PhiB_{R1}$  and 
$\xiB  \equiv \PhiB_{L3} + \PhiB_{R1}$ .
%%%%%%%%%%%%%%%%%%%%%%%%

% ========= SubSection{$(\thC_{ij}, \CPVC)$ Solutions} ============= %
\subsection{$(\thC_{ij}, \CPVC)$ Solutions }
      \label{apdx:NuParam_Solution_AC}
      %\input{z_apdx_NuParam_Sol_2_AC_v6.tex}

% --------- Contents ---------- %
For this case, the rotation matrices,
\begin{eqnarray}
D^L &=& diag \left( e^{i\PhiC_{L1}}, e^{i\PhiC_{L2}}, e^{i\PhiC_{L3}} 
                                               \right) , \nonumber \\
D^R &=& diag \left( 1, e^{i\PhiC_{R2}}, e^{i\PhiC_{R3}} 
                                               \right) , 
\label{eq:Append_C_AC_1}
\end{eqnarray}
are applied to the transformation of $(\theta_{ij}, \CPV)$ from 
representations $\RepA$ to $\RepC$.
The solutions of the nine parameters are listed as follows.
\begin{equation}
\PhiC_{L1} + \PhiC_{R3} =  \CPVA  \hskip0.5cm \rm{and}  \hskip0.5cm
\PhiC_{L3} + \PhiC_{R3} = 0  \; .
\label{eq:Append_C_AC_2}
\end{equation}
\begin{equation}
\sin^2{\PhiC_{L2}} = \frac{ (a~\sin\CPVA )^2}
                          { (a~\sin\CPVA )^2 + (a~\cos\CPVA + b)^2}  \; ,
\label{eq:Append_C_AC_3}
\end{equation}
where $a = \sA_{23} \cA_{12} \sA_{13}$ and $b = \cA_{23} \sA_{12}$ .
\begin{equation}
\sin^2{\PhiC_{25}} = 
      \frac{ (a^{'} \sin\CPVA )^2 }
           { (a^{'} \sin\CPVA )^2 + (a^{'} \cos\CPVA - b^{'})^2 } \; ,
\label{eq:Append_C_AC_4}
\end{equation} 
where $a^{'} = \sA_{23} \sA_{12} \sA_{13}$,  $b^{'} = \cA_{23} \cA_{12}$, 
and $\PhiC_{25} \equiv \PhiC_{L2} + \PhiC_{R2}$ . 
The three mixing angles can thus be obtained in representation $\RepC$. 
\begin{eqnarray}
& & \hskip-1.3cm
  \tan \thC_{12} =
       \frac{\cA_{23}\sA_{12}\cos\PhiC_{L2} 
                + \sA_{23}\cA_{12}\sA_{13}\cos(\CPVA + \PhiC_{L2})}
            {\cA_{23}\cA_{12}\cos\PhiC_{25} 
                - \sA_{23}\sA_{12}\sA_{13}\cos(\CPVA + \PhiC_{25})} ,
\label{eq:Append_C_AC_5}
\end{eqnarray}
\begin{eqnarray}   
& & \hskip-1.3cm
  \cos \thC_{23} = 
       \frac{\cA_{23}\sA_{12}\cos\PhiC_{L2} 
                + \sA_{23}\cA_{12}\sA_{13}\cos(\CPVA + \PhiC_{L2})}
            {\sC_{12}}                                    \nonumber \\ 
& & \hskip-0.2cm 
     = \frac{\cA_{23}\cA_{12}\cos\PhiC_{25} 
                - \sA_{23}\sA_{12}\sA_{13}\cos(\CPVA + \PhiC_{25})}
            {\cC_{12}} ,
\label{eq:Append_C_AC_6}   
\end{eqnarray}
\begin{eqnarray}
& & 
  \sin \thC_{13} = 
       \frac{\sA_{13}} {\cC_{23}}  \; .
\label{eq:Append_C_AC_7}
\end{eqnarray}
The remaining parameters thus can be determined, as shown below.
\begin{eqnarray}
& & \hskip-1.2cm
  \cos \PhiC_{L1} = 
       \frac{( \cC_{12}\cC_{13} )^2 + ( \cA_{12} \cA_{13} )^2 
                                    - ( \sC_{23}\sC_{12}\sC_{13} )^2 }  
            {2 \cC_{12}\cC_{13} \cA_{12}\cA_{13} }.
\label{eq:Append_C_AC_8}
\end{eqnarray}
Therefore, 
$\PhiC_{R3} = \CPVA - \PhiC_{L1}$ and 
$\PhiC_{L3} = -\PhiC_{R3}$ can be determined. 
Finally, the CP-violating phase in representation $\RepC$, $\CPVC$, 
can be obtained using those conditions associated with $\sin \CPVC$ :
\begin{eqnarray}
& & \hskip-1.0cm 
  \sin \CPVC                                \nonumber \\
& & \hskip-1.0cm
  = \frac{ \sA_{23}\cA_{12} \sin \etaC 
          +\cA_{23}\sA_{12}\sA_{13} \sin(\CPVA + \etaC) }
         { \sC_{23}\cC_{12}\cC_{13} } 
                                            \nonumber \\ 
& & \hskip-1.0cm
  = \frac{ \sA_{23}\sA_{12} \sin \PhiC_{L3}
          -\cA_{23}\cA_{12}\sA_{13} \sin(\CPVA + \PhiC_{L3}) }
         { \sC_{23}\sC_{12}\cC_{13} } 
                                            \nonumber \\ 
& & \hskip-1.0cm
  = \frac{-\sA_{12}\cA_{13} \sin(\PhiC_{L1} + \PhiC_{R2}) }  
         { \sC_{23}\cC_{12}\sC_{13} }        \nonumber \\
& & \hskip-1.0cm
  = \frac{-\sA_{23}\cA_{13} \sin(\PhiC_{L2} + \PhiC_{R3}) }  
         { \sC_{23} }                       \nonumber \\
& & \hskip-1.0cm
  = \frac{ \cA_{12}\cA_{13} \sin \PhiC_{L1} }  
         { \sC_{23}\sC_{12}\sC_{13} }  \; ,  
\label{eq:Append_C_AC_9}
\end{eqnarray}
where $\etaC \equiv \PhiC_{L3} + \PhiC_{R2}$ .  
%%%%%%%%%%%%%%%%%%%%%%%%%

% ========== SubSection{$(\thE_{ij}, \CPVE)$ Solutions } ============= %
\subsection{$(\thE_{ij}, \CPVE)$ Solutions }
      \label{apdx:NuParam_Solution_AE}
      %\input{z_apdx_NuParam_Sol_2_AE_v6.tex}

% --------- Contents ----------- %
For this case, the rotation matrices,
\begin{eqnarray}
D^L &=& diag \left( e^{i\PhiE_{L1}}, e^{i\PhiE_{L2}}, e^{i\PhiE_{L3}} 
                                                  \right) , \nonumber \\
D^R &=& diag \left( 1, e^{i\PhiE_{R2}}, e^{i\PhiE_{R3}} 
                                                  \right) , 
\label{eq:Append_C_AE_1}
\end{eqnarray}
are employed to transform $(\thA_{ij}, \CPVA)$ to $(\thE_{ij}, \CPVE)$. 
Below presents the solutions of the nine parameters. 
\begin{equation}
\PhiE_{L1} + \PhiE_{R2} = 0  \hskip0.5cm \rm{and}  \hskip0.5cm
\PhiE_{L3} + \PhiE_{R3} = 0  \; .
\label{eq:Append_C_AE_2}
\end{equation}
\begin{equation}
\sin^2{\PhiE_{L3}} = \frac{ (a~\sin\CPVA )^2}
                           { (a~\sin\CPVA )^2 + (a~\cos\CPVA - b)^2}  \; ,
\label{eq:Append_C_AE_3}
\end{equation}
where $a = \cA_{23} \cA_{12} \sA_{13}$ and $b = \sA_{23} \sA_{12}$ .
\begin{equation}
\sin^2{\PhiE_{25}} = 
      \frac{ (a^{'} \sin\CPVA )^2 }
           { (a^{'} \sin\CPVA )^2 + (a^{'} \cos\CPVA - b^{'})^2 } \; ,
\label{eq:Append_C_AE_4}
\end{equation} 
where $a^{'} = \sA_{23} \sA_{12} \sA_{13}$,  $b^{'} = \cA_{23} \cA_{12}$, 
and $\PhiE_{25} \equiv \PhiE_{L2} + \PhiE_{R2}$ . 
The three mixing angles can thus be acquired in representation $\RepE$. 
\begin{eqnarray}
& & \hskip-1.3cm
  \tan \thE_{12} =
       \frac{ \sA_{12}\cA_{13} }
            { \cA_{23}\cA_{12}\cos\PhiE_{25} 
                - \sA_{23}\sA_{12}\sA_{13}\cos(\CPVA + \PhiE_{25})} ,
\label{eq:Append_C_AE_5}
\end{eqnarray}
\begin{eqnarray}   
& & \hskip-1.3cm
  \cos \thE_{23} = 
       \frac{\sA_{12}\cA_{13} } {\sE_{12}}             \nonumber \\ 
& & \hskip-0.2cm 
     = \frac{\cA_{23}\cA_{12}\cos\PhiE_{25} 
                - \sA_{23}\sA_{12}\sA_{13}\cos(\CPVA + \PhiE_{25})}
            {\cE_{12}} ,
\label{eq:Append_C_AE_6}   
\end{eqnarray}
\begin{eqnarray}
& & \hskip-1.3cm
  \sin \thE_{13} = 
       \frac{ \cA_{23}\cA_{12}\sA_{13}\cos(\CPVA + \PhiE_{L3})
             -\sA_{23}\sA_{12}\cos\PhiE_{L3} }
            {\cE_{23}} .
\label{eq:Append_C_AE_7}
\end{eqnarray}
The remaining parameters thus can be determined, which are shown below.
\begin{eqnarray}
& & \hskip-1.2cm
  \cos \PhiE_{L1} = 
       \frac{( \cE_{12}\cE_{13} )^2 + ( \cA_{12} \cA_{13} )^2 
                                    - ( \sE_{23}\sE_{12}\sE_{13} )^2 }  
            {2 \cE_{12}\cE_{13} \cA_{12}\cA_{13} }.
\label{eq:Append_C_AE_8}
\end{eqnarray}
Therefore, 
$\PhiE_{R2} = - \PhiE_{L1}$, $\PhiE_{L2} = \PhiE_{25}- \PhiE_{R2}$, and 
$\PhiE_{R3} = -\PhiE_{L3}$ can be determined. 
Finally, the CP-violating phase in representation $\RepE$, $\CPVE$, can be 
derived using those conditions associated with $\sin \CPVE$ :
\begin{eqnarray}
& & \hskip-1.0cm 
  \sin \CPVE                                \nonumber \\
& & \hskip-1.0cm
  = \frac{-\sA_{23}\cA_{12} \sin \etaE 
          -\cA_{23}\sA_{12}\sA_{13} \sin(\CPVA + \etaE) }
         {-\sE_{23} } 
                                            \nonumber \\ 
& & \hskip-1.0cm
  = \frac{-\cA_{23}\sA_{12} \sin \PhiE_{L2}
          -\sA_{23}\cA_{12}\sA_{13} \sin(\CPVA + \PhiE_{L2}) }
         { \sE_{23}\cE_{12}\sE_{13} } 
                                            \nonumber \\ 
& & \hskip-1.0cm
  = \frac{\sA_{13} \sin(\CPVA - \Phi^E_{L1} - \Phi^E_{R3}) }  
         {\sE_{23}\sE_{12}\cE_{13} }        \nonumber \\
& & \hskip-1.0cm
  = \frac{-\sA_{23}\cA_{13} \sin(\PhiE_{L2} + \PhiE_{R3}) }  
         { \sE_{23}\cE_{12}\cE_{13} }        \nonumber \\
& & \hskip-1.0cm
  = \frac{ \cA_{12}\cA_{13} \sin \PhiE_{L1} }  
         { \sE_{23}\sE_{12}\sE_{13} }   \; ,  
\label{eq:Append_C_AE_9}
\end{eqnarray}
where $\etaE \equiv \PhiE_{L3} + \PhiE_{R2}$ .  
%%%%%%%%%%%%%%%%%%%%%%

% ========== SubSection{$(\thF_{ij}, \CPVF)$ Solutions} ============ %
\subsection{$(\thF_{ij}, \CPVF)$ Solutions }
      \label{apdx:NuParam_Solution_AF}
      %\input{z_apdx_NuParam_Sol_2_AF_v6.tex}

% -------- Contents ---------- %
For this case, the rotation matrices,
\begin{eqnarray}
D^L &=& diag \left( e^{i\PhiF_{L1}}, e^{i\PhiF_{L2}}, e^{i\PhiF_{L3}} 
                                                  \right) , \nonumber \\
D^R &=& diag \left( e^{i\PhiF_{R1}}, 1, e^{i\PhiF_{R3}} 
                                                  \right) , 
\label{eq:Append_C_AF_1}
\end{eqnarray}
are applied to the transform $(\thA_{ij}, \CPVA)$ to $(\thF_{ij}, \CPVF)$. 
The solutions of the nine parameters are presented as follows.
\begin{equation}
\PhiF_{L1} + \PhiF_{R1} = 0  \hskip0.5cm \rm{and}  \hskip0.5cm
\PhiF_{L3} + \PhiF_{R3} = 0  \; .
\label{eq:Append_C_AF_2}
\end{equation}
\begin{equation}
\sin^2{\PhiF_{24}} = \frac{ (a~\sin\CPVA )^2}
                           { (a~\sin\CPVA )^2 + (a~\cos\CPVA + b)^2}  \; ,
\label{eq:Append_C_AF_3}
\end{equation}
where $a = \sA_{23} \cA_{12} \sA_{13}$, $b = \cA_{23} \sA_{12}$, and
$\PhiF_{24} \equiv \PhiF_{L2} + \PhiF_{R1}$ .
\begin{equation}
\sin^2{\PhiF_{L3}} = 
      \frac{ (a^{'} \sin\CPVA )^2 }
           { (a^{'} \sin\CPVA )^2 + (a^{'} \cos\CPVA + b^{'})^2 } \; ,
\label{eq:Append_C_AF_4}
\end{equation} 
where $a^{'} = \cA_{23} \sA_{12} \sA_{13}$ and  $b^{'} = \sA_{23} \cA_{12}$ . 
The three mixing angles can thus be obtained in representation $\RepF$. 
\begin{eqnarray}
& & \hskip-1.3cm
  \tan \thF_{12} =
       \frac{ \cA_{23}\sA_{12}\cos\PhiF_{24} 
            + \sA_{23}\cA_{12}\sA_{13}\cos(\CPVA + \PhiF_{24})}
            { \cA_{12} \cA_{13} } , 
\label{eq:Append_C_AF_5}
\end{eqnarray}
\begin{eqnarray}   
& & \hskip-1.3cm
  \cos \thF_{13} = 
       \frac{\cA_{12} \cA_{13} } { \cF_{12} }        \nonumber \\
& & \hskip-0.2cm 
     = \frac{ \cA_{23}\sA_{12}\cos\PhiF_{24} 
            + \sA_{23}\cA_{12}\sA_{13}\cos(\CPVA + \PhiF_{24})}
            { \sF_{12} } , 
\label{eq:Append_C_AF_6}   
\end{eqnarray}
\begin{eqnarray}
& & \hskip-1.3cm
  \sin \thF_{23} = 
       \frac{ \sA_{23}\cA_{12}\cos\PhiF_{L3} 
            + \cA_{23}\sA_{12}\sA_{13}\cos(\CPVA + \PhiF_{L3})}
            { \cF_{13} } .
\label{eq:Append_C_AF_7}
\end{eqnarray}
The remaining parameters thus can be determined as shown below.
\begin{eqnarray}   
& & \hskip-1.2cm
  \cos \PhiF_{L1} = 
       \frac{( \cF_{23}\sF_{12} )^2 + ( \sA_{12} \cA_{13} )^2 
                                    - ( \sF_{23}\cF_{12}\sF_{13} )^2 }  
            {2 \cF_{23}\sF_{12} \sA_{12}\cA_{13} }  \; .
\label{eq:Append_C_AF_8}
\end{eqnarray}
Therefore, 
$\PhiF_{R1} = -\PhiF_{L1}$, $\PhiF_{L2} = \PhiF_{24}- \PhiF_{R1}$, and 
$\PhiF_{R3} = -\PhiF_{L3}$ can be determined. 
Finally, the CP-violating phase in representation $\RepF$, $\CPVF$, can be 
obtained using those conditions associated with $\sin \CPVF$ :
\begin{eqnarray}
& & \hskip-1.0cm 
  \sin \CPVF                                \nonumber \\
& & \hskip-1.0cm
  = \frac{-\sA_{23}\sA_{12} \sin \etaF 
          +\cA_{23}\cA_{12}\sA_{13} \sin(\CPVA + \etaF) }
         { \sF_{13} } 
                                            \nonumber \\ 
& & \hskip-1.0cm
  = \frac{-\cA_{23}\cA_{12} \sin \PhiF_{L2}
          +\sA_{23}\sA_{12}\sA_{13} \sin(\CPVA + \PhiF_{L2}) }
         { \sF_{23}\sF_{12}\sF_{13} } 
                                            \nonumber \\ 
& & \hskip-1.0cm
  = \frac{\sA_{13} \sin(\CPVA - \PhiF_{L1} - \PhiF_{R3}) }  
         {\cF_{23}\cF_{12}\sF_{13} }        \nonumber \\
& & \hskip-1.0cm
  = \frac{ \sA_{23}\cA_{13} \sin(\PhiF_{L2} + \PhiF_{R3}) }  
         { \cF_{23}\sF_{12}\sF_{13} }        \nonumber \\
& & \hskip-1.0cm
  = \frac{ \sA_{12}\cA_{13} \sin \PhiF_{L1} }  
         { \sF_{23}\cF_{12}\sF_{13} }   \; ,  
\label{eq:Append_C_AF_9}
\end{eqnarray}
where $\etaF \equiv \PhiF_{L3} + \PhiF_{R1}$ .  
%%%%%%%%%%%%%%%%%%%%%%%%%%

% =========== Section{Neutrino Oscillation Probability} ============= %
\section{Neutrino Oscillation Probability}
     \label{apdx:NuParam_Osci_Prob}
     %\input{z_apdx_Nuparam_OsciProb_v6.tex}

% -------- Contents ---------- %
For oscillation channel $\nu_e \rightarrow \nu_e$, representations $\RepA$ 
and $\RepB$ have simpler forms of survival probabilities:
\begin{eqnarray}
& & \hskip-0.7cm
     \ProbA_{\nu_e \rightarrow \nu_e} =                          \nonumber \\
& & \hskip-0.7cm
     1 - (\cA_{13})^4 \sin^{2}2\thA_{12} \sin^{2}\Delta_{21}  
       - (\cA_{12})^2 \sin^{2}2\thA_{13} \sin^{2}\Delta_{31} \nonumber \\
& & \hskip-0.7cm
       - (\sA_{12})^2 \sin^{2}2\thA_{13} \sin^{2}\Delta_{32}  \; ,
\label{eq:Append_D_11}   
\end{eqnarray}
\begin{eqnarray}
& & \hskip-0.7cm
     \ProbB_{\nu_e \rightarrow \nu_e} =                           \nonumber \\
& & \hskip-0.7cm
     1 - (\cB_{13})^2 \sin^{2}2\thB_{12} \sin^{2}\Delta_{21} 
       - (\cB_{12})^4 \sin^{2}2\thB_{13} \sin^{2}\Delta_{31}  \nonumber \\
& & \hskip-0.7cm
       - (\sB_{13})^2 \sin^{2}2\thB_{12} \sin^{2}\Delta_{32}  \; .
\label{eq:Append_D_12}
\end{eqnarray}
while all other representations have very complicated expressions.
The survival probabilities of $\nu_{\mu} \rightarrow \nu_{\mu}$ in 
representations $\RepC$ and $\RepD$ are of simpler forms than other 
representations:
\begin{eqnarray}
& & \hskip-0.7cm
     \ProbC_{\nu_{\mu} \rightarrow \nu_{\mu}} =                  \nonumber \\
& & \hskip-0.7cm
     1 - (\cC_{23})^4 \sin^{2}2\thC_{12} \sin^{2}\Delta_{21}  
       - (\sC_{12})^2 \sin^{2}2\thC_{23} \sin^{2}\Delta_{31} \nonumber \\
& & \hskip-0.7cm
       - (\cC_{12})^2 \sin^{2}2\thC_{23} \sin^{2}\Delta_{32}  \; ,
\label{eq:Append_D_21}
\end{eqnarray}
\begin{eqnarray}
& & \hskip-0.7cm
     \ProbD_{\nu_{\mu} \rightarrow \nu_{\mu}} =                  \nonumber \\
& & \hskip-0.7cm
     1 - (\cD_{23})^2 \sin^{2}2\thD_{12} \sin^{2}\Delta_{21}  
       - (\sD_{23})^2 \sin^{2}2\thD_{12} \sin^{2}\Delta_{31} \nonumber \\
& & \hskip-0.7cm
       - (\cD_{12})^4 \sin^{2}2\thD_{23} \sin^{2}\Delta_{32}  \; .
\label{eq:Append_D_22}
\end{eqnarray}
%%%%%%%%%
Furthermore, representations $\RepE$ and $\RepF$ have simpler forms for 
survival probabilities of $\nu_{\tau} \rightarrow \nu_{\tau}$:
\begin{eqnarray}
& & \hskip-0.7cm
     \ProbE_{\nu_{\tau} \rightarrow \nu_{\tau}} =                \nonumber \\
& & \hskip-0.7cm
     1 - (\sE_{13})^2 \sin^{2}2\thE_{23} \sin^{2}\Delta_{21}  
       - (\cE_{23})^4 \sin^{2}2\thE_{13} \sin^{2}\Delta_{31} \nonumber \\
& & \hskip-0.7cm
       - (\cE_{13})^2 \sin^{2}2\thE_{23} \sin^{2}\Delta_{32}  \; ,
\label{eq:Append_D_31}
\end{eqnarray}
\begin{eqnarray}
& & \hskip-0.7cm
     \ProbF_{\nu_{\tau} \rightarrow \nu_{\tau}} =                \nonumber \\
& & \hskip-0.7cm
     1 - (\sF_{23})^2 \sin^{2}2\thF_{13} \sin^{2}\Delta_{21}  
       - (\cF_{23})^2 \sin^{2}2\thF_{13} \sin^{2}\Delta_{31}  \nonumber \\
& & \hskip-0.7cm
       - (\cF_{13})^4 \sin^{2}2\thF_{23} \sin^{2}\Delta_{32}  \; .
\label{eq:Append_D_32}
\end{eqnarray}
%%%%%%%%%%%%%%%%%%%%%%%

The oscillation channel $\nu_{\mu} \rightarrow \nu_e$ does not have a simpler 
form if compared to the survival probabilities of $\nu_{\alpha} \rightarrow
\nu_{\alpha}$. One, however, still can work out some simpler expressions 
in representations $\RepA$--$\RepD$ for the limit of 
$\Delta_{31} \sim \Delta_{32}$, which are
\begin{eqnarray}
& & \hskip-1.0cm 
      \ProbA_{\nu_{\mu} \rightarrow \nu_e} =                     \nonumber \\
& & \hskip-1.0cm
      (\sA_{23})^2 \sin^{2}2\thA_{13} \sin^{2}\Delta_{31} +
      2 (\cA_{13})^2 \sin 2\thA_{12} \cdot                   \nonumber \\
& & \hskip-1.0cm
      ( \cA_{23} \cA_{12} - \sA_{23} \sA_{12} \sA_{13} ) 
      ( \cA_{23} \sA_{12} + \sA_{23} \cA_{12} \sA_{13} ) 
      \sin^{2}\Delta_{21}                                   \; , 
\label{eq:Append_D_41}
\end{eqnarray}
\begin{eqnarray}
& & \hskip-1.5cm
      \ProbB_{\nu_{\mu} \rightarrow \nu_e} =                      \nonumber \\
& & \hskip-1.5cm
      \left[ (\sB_{13})^2 \cB_{23} \sin 2\thB_{12} - 
             \sB_{23} \cB_{12} \sin 2\thB_{13} \right]^2 \sin^{2}\Delta_{31}
                                                                 \nonumber \\
& & \hskip-1.5cm
      + \left[ (\cB_{23})^2 (\cB_{13})^2 
             \sin^{2}2\thB_{12}        \right.                \nonumber \\
& & \hskip-1.5cm
    + \left. \frac{1}{2} \cB_{12} \sin 2\thB_{23} \sin 2\thB_{12}
             \sin 2\thB_{13}  \right] \sin^{2}\Delta_{21} \; ,
\label{eq:Append_D_42}
\end{eqnarray}
\begin{eqnarray}
& & \hskip-1.0cm 
      \ProbC_{\nu_{\mu} \rightarrow \nu_e} =                     \nonumber \\
& & \hskip-1.0cm
      (\sC_{13})^2 \sin^{2}2\thC_{23} \sin^{2}\Delta_{31} +
      2 (\cC_{23})^2 \sin 2\thC_{12} \cdot                   \nonumber \\
& & \hskip-1.0cm
      ( \sC_{12} \cC_{13} - \sC_{23} \cC_{12} \sC_{13} ) 
      ( \cC_{12} \cC_{13} + \sC_{23} \sC_{12} \sC_{13} ) 
      \sin^{2}\Delta_{21}                                   \; , 
\label{eq:Append_D_43}
\end{eqnarray}
\begin{eqnarray}
& & \hskip-1.5cm
      \ProbD_{\nu_{\mu} \rightarrow \nu_e} =                      \nonumber \\
& & \hskip-1.5cm
      \left[ (\sD_{23})^2 \cD_{13} \sin 2\thD_{12} + 
             \cD_{12} \sD_{13} \sin 2\thD_{23} \right]^2 \sin^{2}\Delta_{31}
                                                                 \nonumber \\
& & \hskip-1.5cm
     + \left[ (\cD_{23})^2 (\cD_{13})^2 
              \sin^{2}2\thD_{12} \right.                      \nonumber \\ 
& & \hskip-1.5cm  
    - \left. \frac{1}{2} \cD_{12} \sin 2\thD_{23} \sin 2\thD_{12}
             \sin 2\thD_{13}  \right] \sin^{2}\Delta_{21}  \; . 
\label{eq:Append_D_44}
\end{eqnarray}
%%%%%%%%%%%%%%%%%%%%%%%%%%

The oscillation channel $\nu_{\mu} \rightarrow \nu_{\tau}$, on the other 
hand, has simpler forms in representations $\RepC$--$\RepF$ in the limit of 
$\Delta_{31} \sim \Delta_{32}$, which are
\begin{eqnarray}
& & \hskip-1.0cm 
      \ProbC_{\nu_{\mu} \rightarrow \nu_{\tau}} =                \nonumber \\
& & \hskip-1.0cm
      (\cC_{13})^2 \sin^{2}2\thC_{23} \sin^{2}\Delta_{31} +
      2 (\cC_{23})^2 \sin 2\thC_{12} \cdot                   \nonumber \\
& & \hskip-1.0cm
      ( \sC_{12} \sC_{13} + \sC_{23} \cC_{12} \cC_{13} ) 
      ( \cC_{12} \sC_{13} - \sC_{23} \sC_{12} \cC_{13} ) 
      sin^{2}\Delta_{21}   \; , 
\label{eq:Append_D_51}  
\end{eqnarray}
\begin{eqnarray}
& & \hskip-1.5cm
      \ProbD_{\nu_{\mu} \rightarrow \nu_{\tau}} =                      \nonumber \\
& & \hskip-1.5cm
      \left[ (\sD_{23})^2 \sD_{13} \sin 2\thD_{12} - 
             \cD_{12} \cD_{13} \sin 2\thD_{23} \right]^2 \sin^{2}\Delta_{31}
                                                                 \nonumber \\
& & \hskip-1.5cm
     + \left[ (\cD_{23})^2 (\sD_{13})^2 
              \sin^{2}2\thD_{12} \right.                      \nonumber \\ 
& & \hskip-1.5cm  
    + \left. \frac{1}{2} \cD_{12} \sin 2\thD_{23} \sin 2\thD_{12}
             \sin 2\thD_{13}  \right] \sin^{2}\Delta_{21}  \; . 
\label{eq:Append_D_52}
\end{eqnarray}
\begin{eqnarray}
& & \hskip-1.5cm
      \ProbE_{\nu_{\mu} \rightarrow \nu_{\tau}} =                      \nonumber \\
& & \hskip-1.5cm
      \left[ (\cE_{13})^2 \cE_{12} \sin 2\thE_{23} - 
              \cE_{23} \sE_{12} \sin 2\thE_{13} \right]^2 \sin^{2}\Delta_{31}
                                                                 \nonumber \\
& & \hskip-1.5cm
     + \left[ (\cE_{12})^2 (\sE_{13})^2 
              \sin^{2}2\thE_{23} \right.                      \nonumber \\ 
& & \hskip-1.5cm  
    + \left. \frac{1}{2} \cE_{23} \sin 2\thE_{23} \sin 2\thE_{12}
             \sin 2\thE_{13}  \right] \sin^{2}\Delta_{21}  \; . 
\label{eq:Append_D_53}
\end{eqnarray}
\begin{eqnarray}
& & \hskip-1.5cm
      \ProbF_{\nu_{\mu} \rightarrow \nu_{\tau}} =                      \nonumber \\
& & \hskip-1.5cm
      \left[ (\cF_{23})^2 \sF_{12} \sin 2\thF_{13} - 
             \cF_{12} \cF_{13} \sin 2\thF_{23} \right]^2 \sin^{2}\Delta_{31}
                                                                 \nonumber \\
& & \hskip-1.5cm
     + \left[ (\sF_{23})^2 (\sF_{12})^2 
              \sin^{2}2\thF_{13} \right.                      \nonumber \\ 
& & \hskip-1.5cm  
    + \left. \frac{1}{2} \cF_{13} \sin 2\thF_{23} \sin 2\thF_{12}
             \sin 2\thF_{13}  \right] \sin^{2}\Delta_{21}  \; . 
\label{eq:Append_D_54}
\end{eqnarray}
%%%%%%%%%%%%%%%%%%%%%%%%%%%

% ======================= References =============================== % 
%\input{z_ref_NuParam_v6.tex}

% =================== End of Document ================================ %
\end{document}